\documentclass[]{aa}
\usepackage[nolist]{acronym}
\usepackage[displaymath, mathlines]{lineno}
\usepackage{graphicx}
\usepackage{txfonts}
\usepackage{natbib,twoopt}
\bibpunct{(}{)}{;}{a}{}{,} 
\usepackage{hyperref}
\usepackage{nameref}
\usepackage{cleveref}
\crefname{figure}{Fig.}{Figs.}
\Crefname{figure}{Fig.}{Figs.}
\usepackage{nccmath}
\usepackage{subcaption}
\usepackage{siunitx}
\usepackage{booktabs} 
\usepackage{multirow} 
\usepackage{threeparttable} 
\usepackage{tabularx} 
\usepackage{array} 
\usepackage{bm} 
\usepackage[table]{xcolor} 

\usepackage{scalerel}
\usepackage{tikz}
\usetikzlibrary{svg.path}

\definecolor{orcidlogocol}{HTML}{A6CE39}
\tikzset{
  orcidlogo/.pic={
    \fill[orcidlogocol] svg{M256,128c0,70.7-57.3,128-128,128C57.3,256,0,198.7,0,128C0,57.3,57.3,0,128,0C198.7,0,256,57.3,256,128z};
    \fill[white] svg{M86.3,186.2H70.9V79.1h15.4v48.4V186.2z}
                 svg{M108.9,79.1h41.6c39.6,0,57,28.3,57,53.6c0,27.5-21.5,53.6-56.8,53.6h-41.8V79.1z M124.3,172.4h24.5c34.9,0,42.9-26.5,42.9-39.7c0-21.5-13.7-39.7-43.7-39.7h-23.7V172.4z}
                 svg{M88.7,56.8c0,5.5-4.5,10.1-10.1,10.1c-5.6,0-10.1-4.6-10.1-10.1c0-5.6,4.5-10.1,10.1-10.1C84.2,46.7,88.7,51.3,88.7,56.8z};
  }
}

\newcommand\orcidicon[1]{\href{https://orcid.org/#1}{\mbox{\scalerel*{
\begin{tikzpicture}[yscale=-1,transform shape]
\pic{orcidlogo};
\end{tikzpicture}
}{|}}}}

\newcommand{\orcididDario}{0000-0001-9282-9462}
\newcommand{\orcididIgnas}{0000-0003-1624-3667}
\newcommand{\orcididSam}{0000-0003-4760-6168}
\newcommand{\orcidNatalie}{0009-0009-6634-1741}
\newcommand{\orcididSid}{0000-0001-9552-3709}

\definecolor{cobalt}{rgb}{0.06, 0.2, 0.65}
\hypersetup{
  colorlinks,
  citecolor=cobalt,
  linkcolor=[rgb]{0.8, 0.2, 1.0},
  urlcolor=cobalt,
}

\definecolor{lightgray}{gray}{0.95}

\makeatletter
\renewcommand*\aa@pageof{, page \thepage{} of \pageref*{LastPage}}

\newcommand{\logg}{\ensuremath{\log g}}

\def\kms{\ifmmode{\rm km\th s^{-1}}\else km\th s$^{-1}$\fi}
\def\th{\thinspace}
\newcommand{\Mjup}{M$_{\text{Jup}}$}
\newcommand{\Rjup}{R$_{\text{Jup}}$}

\newcommand{\twelveCO}{\textsuperscript{12}CO}
\newcommand{\thirteenCO}{\textsuperscript{13}CO}
\newcommand{\CeighteenO}{C\textsuperscript{18}O}
\newcommand{\CseventeenO}{C\textsuperscript{17}O}
\newcommand{\fifteenammonia}{\textsuperscript{15}NH\textsubscript{3}}

\newcommand{\ratio}[3]{\textsuperscript{#2}#1/\textsuperscript{#3}#1}

\newcommand{\pRT}{\texttt{petitRADTRANS}}
\newcommand{\micron}{$\mu$m}

\newcommand{\Cratio}{\textsuperscript{12}C/\textsuperscript{13}C}
\newcommand{\Oratio}{\textsuperscript{16}O/\textsuperscript{18}O}

\newcommand{\water}{H$_2$O}
\newcommand{\methane}{CH$_4$}
\newcommand{\COtwo}{CO$_2$}

\newcommand{\eighteenOwater}{H\textsubscript{2}\textsuperscript{18}O}
\newcommand{\CRIRES}{CRIRES$^+$}

\newcommand{\leiden}{Leiden Observatory, Leiden University, P.O. Box 9513, 2300 RA, Leiden, The Netherlands}
\newcommand{\warwick}{Department of Physics, University of Warwick, Coventry CV4 7AL, UK}
\newcommand{\warwickexoplanets}{Centre for Exoplanets and Habitability, University of Warwick, Gibbet Hill Road, Coventry CV4 7AL, UK}

\newcommandtwoopt{\citeads}[3][][]{\href{http://adsabs.harvard.edu/abs/#3}
{\def\hyper@linkstart##1##2{}
    \let\hyper@linkend\@empty\citealp[#1][#2]{#3}}}
\newcommandtwoopt{\citepads}[3][][]{\href{http://adsabs.harvard.edu/abs/#3}
{\def\hyper@linkstart##1##2{}
    \let\hyper@linkend\@empty\citep[#1][#2]{#3}}}

\makeatother

\begin{document} 
\title{Disentangling disc and atmospheric signatures of young brown dwarfs with JWST/NIRSpec}

\titlerunning{Disentangling disc and atmospheric signatures of young brown dwarfs with JWST/NIRSpec}
\author{D. Gonz\'{a}lez Picos       \inst{\ref{leiden}\orcidicon{\orcididDario}}
           \and S. de Regt          \inst{\ref{leiden}\orcidicon{\orcididSam}}
           \and S. Gandhi           \inst{\ref{warwick},\ref{warwickexoplanets}\orcidicon{\orcididSid}}
           \and N. Grasser          \inst{\ref{leiden}\orcidicon{\orcidNatalie}}
           \and I.A.G. Snellen      \inst{\ref{leiden}\orcidicon{\orcididIgnas}}}

\institute{
    \leiden\\ \email{picos@strw.leidenuniv.nl}\label{leiden}
    \and \warwick\label{warwick}
    \and \warwickexoplanets\label{warwickexoplanets}
}

   \date{Received 2025-05-26; accepted 2025-09-11}

\abstract
{Young brown dwarfs serve as analogues of giant planets and provide benchmarks for atmospheric and formation models. JWST has enabled access to near-infrared spectra of brown dwarfs with unprecedented sensitivity.}
  {We aim to constrain the chemical compositions, temperature structures, isotopic ratios, and the properties of the continuum and line emission from their circumstellar discs.}
  {We perform atmospheric retrievals and disc modelling on JWST/NIRSpec medium-resolution ($R\sim2700$) spectra covering 0.97--5.27~\micron{}. Our approach combines radiative transfer, line-by-line opacities, parameterised temperature profiles, and flexible equilibrium chemistry for the atmospheres. We also include a ring component from the disc, with blackbody continuum and optically thin CO slab emission.}
  {We detect and constrain more than twenty molecular and atomic species in the atmospheres, including \twelveCO{}, \water{}, \COtwo{}, SiO, and several hydrides. The CO fundamental band at 4.6~\micron{} enables detections of \thirteenCO{} and \CeighteenO{}. We report isotope ratios of carbon: $^{12}$C/$^{13}$C$ = 79^{+14}_{-11}$ (TWA 27A) and 75$^{+2}_{-2}$ (TWA 28), and oxygen: $^{16}$O/$^{18}$O$ = 645^{+80}_{-70}$ (TWA 27A) and 681$^{+53}_{-50}$ (TWA 28) from water isotopologues. Both objects show significant excess infrared emission, which we model as warm ($\approx$650~K) blackbody rings. We identify optically thin CO emission from hot gas ($\geq 1600$~K) in the discs, necessary to reproduce the redder part of the spectra. The atmospheric carbon-to-oxygen ratios are 0.54$\pm$0.02 (TWA 27A) and 0.59$\pm$0.02 (TWA 28), consistent with solar values.}
  {We characterise the atmospheres and discs of two young brown dwarfs through simultaneous constraints on temperature, composition, isotope ratios, and disc properties. These observations demonstrate the ability of JWST/NIRSpec to study young objects, enabling future studies of circumplanetary discs.}

  
\maketitle
\section{Introduction}\label{sec:introduction}
Young brown dwarfs (YBDs) share chemical compositions and temperatures with hot gas giant exoplanets. Atmospheric characterisation of YBDs serve as a benchmark for the study of directly imaged companions such as TWA 27B, GQ Lup B, and the HR 8799 planets \citep{chauvinGiantPlanetCompanion2005,neuhaeuserEvidenceComovingSubstellar2005,maroisDirectImagingMultiple2008}. The presence of numerous molecular features and atomic lines complicates the modelling of substellar atmospheres \citep{kirkpatrickUniqueSpectrumBrown1993,allardModelAtmospheresSubDwarf1995} but advances in linelists and retrieval methods are bridging the gap between observations and models. Evolutionary models require detailed opacity information to calculate cooling rates and predict the spectral energy distribution for a range of masses and ages \citep{baraffeEvolutionaryModelsLowmass2002,fortneyYoungJupitersAre2005}. Condensates and clouds can strongly affect observed spectra, especially as objects cool through the L- and T-dwarf sequence \citep{ackermanPrecipitatingCondensationClouds2001,saumonEvolutionDwarfsColor2008,hellingAtmospheresBrownDwarfs2014}. At $T \gtrsim 2000$~K, clouds are less likely and photospheres are expected to be cloud-free \citep{tremblinFINGERINGCONVECTIONCLOUDLESS2015,marleySonoraBrownDwarf2021}. Accretion can be relevant for young systems and may imprint strong emission lines in the optical and near-infrared (e.g., H$\alpha$, Pa$\beta$, Br$\gamma$) \citep{muzerolleMeasuringAccretionYoung2005,nattaAccretionBrownDwarfs2004,bowlerDISKPLANETARYMASSCOMPANION2011,alcalaXshooterSpectroscopyYoung2014}. YBDs are thought to undergo an accretion phase and show disc fractions similar to young stars \citep{luhmanFormationEarlyEvolution2012}. Discs are inferred from infrared excess and veiling of photospheric lines \citep{hartiganHowUnveilTauri1989,mcclureCHARACTERIZINGSTELLARPHOTOSPHERES2013,christiaensEvidenceCircumplanetaryDisk2019,sanchisDemographicsDisksYoung2020}. At longer wavelengths, disc emission can dominate over the photospheric signal and interfere with atmospheric features \citep{obergObservingCircumplanetaryDisks2023}.

Atmospheric composition, including elemental and isotopic ratios, provides insight into the formation and evolution of brown dwarfs and exoplanets \citep{obergEFFECTSSNOWLINESPLANETARY2011,molliereInterpretingAtmosphericComposition2022}. Isotopic ratios are thought to be tracers of formation environments \citep{molliereDetectingIsotopologuesExoplanet2019,zhang13COrichAtmosphereYoung2021}, but are challenging to measure due to weak features and blending. Ground-based high-resolution spectroscopy (HRS) can constrain \Cratio{} using high signal-to-noise \textit{K}-band data (2.3~\micron{}) \citep[e.g.][]{zhang13COrichAtmosphereYoung2021,zhang12CO132021,xuanValidationElementalIsotopic2024,gonzalezpicosChemicalEvolutionImprints2025}. However, the strongest CO features are in the fundamental band at 4.6~\micron{}, which is difficult to observe from the ground. \cite{crossfieldUnusualIsotopicAbundances2019} detected \thirteenCO{} and \CeighteenO{} in M dwarfs using high-resolution \textit{M}- and \textit{K}-band spectroscopy, finding ratios higher than solar. Access to the fundamental CO band at 4.6~\micron{} is essential for robust isotopic measurements of carbon and oxygen, and for faint sources may be the only suitable approach.

Space-based infrared spectroscopy with JWST/NIRSpec and MIRI enables the detection of individual isotopologues in cool atmospheres \citep{milesJWSTEarlyreleaseScience2023,fahertyMethaneEmissionCool2024,billerJWSTWeatherReport2024}. These instruments provide high sensitivity for characterising low-mass atmospheres and allow detailed measurements of composition, temperature structure, and clouds. Recent JWST/NIRSpec results include detections of \thirteenCO{}, \CeighteenO{}, and \CseventeenO{} in VHS 1256 b \citep{gandhiJWSTMeasurements13C2023a}, \thirteenCO{} in WISE J1828 \citep{lewHighprecisionAtmosphericCharacterization2024} and 2MASS 04150935 \citep{hoodHighPrecisionAtmosphericConstraints2024}, and \fifteenammonia{} in WISE J1828 and WISE 0855 using MIRI/MRS \citep{barrado15NH3AtmosphereCool2023,kuhnleWaterDepletion15NH32024}. JWST/NIRSpec also enabled the measurement of the deuterium-to-hydrogen ratio via CH$_3$D \citep{rowlandProtosolarDtoHAbundance2024}. Analysis of a T-dwarf binary by \citet{matthewsHCNC2H2Atmosphere2025} reported the unexpected presence of C$_2$H$_2$ and HCN as potential opacity sources in their atmospheres. JWST observations are pushing the requirements of atmospheric models to accurately describe the numerous spectral features and diversity observed in the atmospheres of brown dwarfs and exoplanets \citep[e.g.][]{fahertyMethaneEmissionCool2024}.

In this work, we present atmospheric retrievals of the young brown dwarfs TWA 27A and TWA 28, using the full JWST/NIRSpec coverage in the high-resolution mode (M24; \citealt{manjavacasMediumresolution09753Mm2024}) to constrain temperature structure, molecular abundances, isotopic ratios, and disc properties. TWA 27A and TWA 28 are members of the TW Hydrae association (TWA; \citealt{delarezaDiscoveryNewIsolated1989,kastnerXrayMolecularEmission1997,webbDiscoverySevenTauri1999}), the nearest young moving group ($<100$~pc, $10 \pm 3$~Myr; \citealt{bellSelfconsistentAbsoluteIsochronal2015}). TWA is a benchmark for young associations, with a well-characterised initial mass function and detailed kinematic studies (e.g., \citealt{gagneBANYANIXInitial2017,luhmanCensusTWHya2023}).

TWA 27A is a $21\pm6$~\Mjup{} YBD with $T_\mathrm{eff} \approx 2400-2600~$K, $\log g = 4.0\pm0.5$, and spectral type M8$\beta$ \citep{gizisBrownDwarfsTW2002,mamajekMovingClusterDistance2005,venutiXshooterSpectroscopyYoung2019,cooperUltracoolSpectroscopicOutliers2024}. Its companion, TWA 27B (2M1207~b), was the first directly imaged planetary-mass object \citep{chauvinGiantPlanetCompanion2005}. Previous studies focused on the companion's mass and fundamental properties (e.g., \citealt{mamajekImprobableSolutionUnderluminosity2007,ricciALMAObservationsYoung2017,luhmanJWSTNIRSpecObservations2023}) and recent analysis of JWST/NIRSpec presented a comprehensive analysis of its atmospheric parameters and reported the presence of a patchy clouds \citep{zhangELementalAbundancesPlanets2025}.

TWA 28 (SS1102) is similar, with $20\pm5$~\Mjup{}, $T_\mathrm{eff} \approx 2400-2600~$K, $\log g = 4.0\pm0.5$, and spectral type M8.5$\beta$ \citep{scholzSSSPMJ11023431Probable2005,venutiXshooterSpectroscopyYoung2019,cooperUltracoolSpectroscopicOutliers2024}. Observations confirm both objects as young, low-mass sources with circumstellar discs \citep{riazNewBrownDwarf2008,morrowObservationsDisksBrown2008,herczegMEASURINGTINYMASS2009,venutiXshooterSpectroscopyYoung2019}. Spitzer detected infrared excesses and dust settling \citep{morrowObservationsDisksBrown2008}. JWST/NIRSpec spectra require an additional emitting source to model flux beyond 2.5~\micron{} \citep[M24]{manjavacasMediumresolution09753Mm2024}.

In Section~\ref{sec:methods}, we describe our data reduction and retrieval approach. Section~\ref{sec:results} presents the retrieval results, while Section~\ref{sec:discussion} interprets these findings and compares them with previous work. Section~\ref{sec:conclusions} summarises the main results.

\section{Methods}\label{sec:methods}

\subsection{Observations}\label{sec:observations}
Observations of TWA 28 and TWA 27A were conducted using the integral field unit (IFU) of the JWST (\citealt{bokerInfraredSpectrographNIRSpec2022,gardnerJamesWebbSpace2023}) as part of the Guaranteed Time Observations (GTO) programme 1270 (PI: S. Birkmann). The observations covered a wavelength range of 0.97--5.27~\micron{}, utilising the highest resolution mode of the Near-Infrared Spectrograph (NIRSpec) with the three gratings G140H, G235H, and G395H (see M24 for more details on the observing setup).

We obtained stage 2 data (software version 1.18.0) products from the Barbara A. Mikulski Archive for Space Telescopes (MAST). We performed a custom extraction routine to obtain spectra from the calibrated images at each dithering position using a circular aperture of four pixels (0.4 arcsec). We found that this radius optimises the signal-to-noise ratio and minimises fringing effects known to be present in the IFU \citep{law3DDrizzleAlgorithm2023}. Our extraction leads to fewer bad pixels than M24, likely due to pipeline improvements and our custom routine to reject outliers from spectra at different dithering positions. We applied a wavelength-dependent aperture correction by comparing extractions with a six-pixel aperture, increasing the flux by five to ten per cent towards redder wavelengths. Absolute flux calibration is not critical for our analysis and we discuss potential pitfalls in Section~\ref{sec:discussion}.

\subsection{Atmospheric modelling}\label{sec:atmospheric_model}
We generated atmospheric models using the radiative transfer code \pRT{} (v2.7, \citealt{mollierePetitRADTRANSPythonRadiative2019}), which computes synthetic spectra from line-by-line and continuum opacities, atmospheric temperature, volume mixing ratios of chemical species, and surface gravity.

\subsubsection{Opacities}\label{sec:opacities}
We used state-of-the-art line lists to generate line-by-line high-resolution opacities. Opacities were calculated with \texttt{pyROX}\footnote{\url{https://github.com/samderegt/pyROX}} \citep{regtpyROX} using transitions, states, and partition functions from ExoMol \citep{tennyson2024ReleaseExoMol2024}, HITEMP/HITRAN \citep{rothmanHITEMPHightemperatureMolecular2010,gordonHITRAN2016MolecularSpectroscopic2017}, and Kurucz \citep{castelliNewGridsATLAS92003}. \Cref{tab:opacity_references} lists all opacity sources, including line lists, Rayleigh scattering, and quasi-continuum opacities.

\subsubsection{Temperature profile}\label{sec:temperature_profile}
The temperature of each atmospheric layer is set following the parameterisation described in \citealt{gonzalezpicosESOSupJupSurvey2025}, which extends the gradient-based temperature profile of \cite{zhangELementalAbundancesPlanets2023a}. We define seven pressure levels where the temperature gradients are measured, fitting for the location of these pressure levels and the values of the temperature gradients. The temperature gradient at a finer grid of equally spaced forty atmospheric layers is determined by linear interpolation of the temperature gradients at the seven pressure levels. This approach explores a wide range of temperature profiles, balancing physically motivated prior knowledge with the flexibility to fit the data. Our prior distributions ensure a decreasing temperature with decreasing pressure, i.e., positive temperature gradients (see \Cref{tab:free_params_comparison_0}).

\subsubsection{Chemical composition}\label{sec:chemical_composition}

We determine the abundances of chemical species using an extended chemical equilibrium approach. For each species, we fit deviations from equilibrium, creating a hybrid model that combines the flexibility of free-composition retrievals \citep[e.g.][]{lineUniformAtmosphericRetrieval2015} with the physically motivated structure of chemical equilibrium models \citep[e.g.][]{marleySonoraBrownDwarf2021}. Individual species abundances can vary independently, whilst leveraging altitude-dependent profiles and priors informed by chemical equilibrium.

A grid of chemical abundances as a function of pressure and temperature is generated using the software package \texttt{FastChem} (v2.0, \citealt{kitzmannFastchemCondEquilibrium2023}). At each model evaluation, the chemical composition for the given temperature profile is calculated via linear grid interpolation. Offset parameters are then applied to the nominal abundances. To assess the impact of individual species and achieve a better fit to the data, we allow the volume mixing ratio of individual species ($X_i$) to deviate from chemical equilibrium by a corresponding factor $\alpha_i$ in log-space,
\begin{equation}\label{eq:alpha}
    \log X_i = \log X_{i, \text{eq}} + \alpha_i.
\end{equation}
The $\alpha$-parameters are sampled from a normal distribution with a mean of 0 and a standard deviation of 1, enabling the exploration of non-equilibrium conditions while favoring chemical equilibrium in the absence of strong model preference.

\subsection{Slab model}\label{sec:slab_model}
We include a disc component in the atmospheric model that consists of a blackbody continuum $B_{\lambda}(T_{\rm BB})$ and line emission $L_{\lambda}$. We introduce two free parameters to describe the blackbody emission, the temperature $T_{\rm BB}$ and the radius $R_{\rm BB}$ of the emitting region. The prior ranges are set to uniformly cover a wide parameter space, including the values found by the literature \citep{boucherBANYANVIIINew2016}. 

We consider additional line emission from optically-thin gas in the disc by including a slab model with a single excitation temperature $T^{\rm slab}_{\rm ex}$, gas column density $N^{\rm slab}_{\rm col}$ and the extent of the ring as defined by the effective radius $R^{\rm slab}$, corresponding to the surface of an annulus extending from inner to outer edge of the ring. A precomputed grid of slab models is calculated using the software package \texttt{iris} \citep{munoz-romeroJWSTMIRISpectroscopyWarm2024} spanning a range of excitation temperatures from 300 to 2000~K in steps of 50~K and gas column densities from $10^{15}$ to $10^{20}$~cm$^{-2}$ with a log-spacing of approximately 0.5~dex. We include three species in the slab model, \twelveCO{}, \thirteenCO{}, and \water. Initially we attempted to retrieve the slab parameters for each species but found that only \twelveCO{} was well constrained by the data. We therefore only fit for the parameters of the slab model for \twelveCO{} but include the other species in the model. The column density of \thirteenCO{} is determined from $N_{\rm slab}$(\twelveCO{}) and the atmospheric isotope ratio \Cratio{}, assuming the atmospheric \Cratio{} is representative of the disc. The column density is fixed to that of \twelveCO{} in the slab model. We note that the inclusion of \thirteenCO{} and \water{} in the slab model does not noticeably improve the fit to the data but is included to provide a more complete model of the disc and avoid potential biases in the isotopic ratios.

\begin{figure*}[h!]
    \centering
    \includegraphics[width=\textwidth]{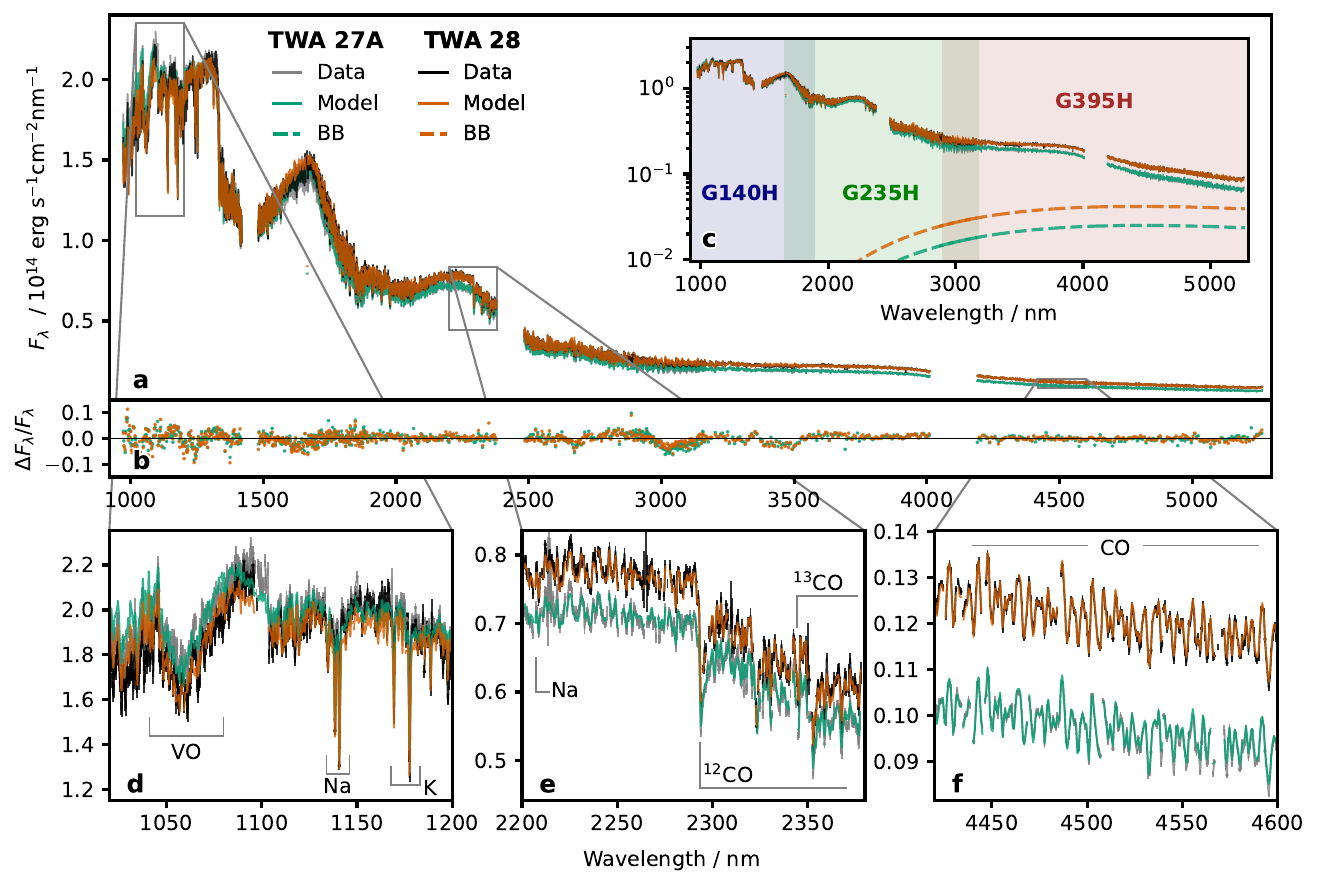}
    \caption{JWST/NIRSpec observations and atmospheric models. Full wavelength coverage of JWST/NIRSpec observations and model fits for TWA 27A and TWA 28. \textbf{a} shows observed spectra (black) and best-fit model spectra (coloured lines) for TWA 27A (blue) and TWA 28 (orange) across the full wavelength range. \textbf{b} displays relative residuals $\Delta F_{\lambda}/F_{\lambda}= (F_{\lambda,{\rm obs}}-F_{\lambda,{\rm model}})/F_{\lambda,{\rm obs}}$ of the best-fit model. \textbf{c} presents the same data in log-scale, with blackbody contribution indicated and coverage of each grating (G140H, G235H, G395H) shown as shaded regions. The increasing offset between datasets at redder wavelengths reflects distinct blackbody contributions in each system. \textbf{d}, \textbf{e}, and \textbf{f} show zoomed regions from each grating with key absorption features labelled.}
\label{fig:full_spec}
\end{figure*}

\subsection{Instrumental line profile}\label{sec:instrumental_line_profile}
The model spectra are convolved with a Gaussian instrumental line profile with the instrumental full width at half maximum (FWHM) at each wavelength 
\footnote{Spectral resolution curves are available at \url{https://jwst-docs.stsci.edu/}.}. Additional broadening effects such as rotational broadening are not included since they are expected to be significantly smaller than the instrumental broadening.

\subsection{Bayesian retrieval}\label{sec:bayesian_retrieval}
We explore the high-dimensional parameter space of the atmospheric and disc models using a Bayesian inference approach. Our models contain up to 51 free parameters (see priors and descriptions in \Cref{tab:free_params_comparison_0}). We use the nested sampling algorithm \citep{ferozMultiNestEfficientRobust2009} as implemented in the \texttt{PyMultiNest} package \citep{buchnerPyMultiNestPythonInterface2016}. We ran retrievals with importance nested sampling \citep{ferozImportanceNestedSampling2019} using 800 live points in constant efficiency mode of 5\% and stopping criterion at Bayesian evidence $\Delta\ln\mathcal{Z} = 0.5$. We assessed the convergence of the results with different numbers of live points and determined 800 to be a reasonable number to balance stable results and computational cost. We note that the Bayesian evidence in constant efficiency mode may be inaccurate \citep{ferozImportanceNestedSampling2019}; however, we do not employ this quantity for model comparison or other purposes. While 51 parameters approaches the upper limit for nested sampling, our convergence tests with different live points suggest the results are stable. Retrieval studies of similar objects (e.g. \cite{gandhiJWSTMeasurements13C2023a,hoodHighPrecisionAtmosphericConstraints2024}) are increasing in complexity and thus elevating the number of free parameters. Our model requires additional parameters for the disc component and detailed temperature structure that push the dimensionality of the parameter space over 40, akin to the recent analysis of a young, cloudy brown dwarf with JWST/NIRSpec \citep{molliereEvidenceSiOCloud2025}. Novel methods for high-dimensional inference may be necessary to keep up with the increasing complexity of these models (e.g. \citep{vasistNeuralPosteriorEstimation2023,gebhardFlowMatchingAtmospheric2025}).

We assume a Gaussian likelihood function for the data, which can be expressed as log-likelihood:

\begin{equation}
\ln \mathcal{L}(\theta) = -\frac{1}{2}N\ln(2\pi) - \frac{1}{2}\ln|\Sigma| - \frac{1}{2} (d - m(\theta))^T \Sigma^{-1} (d - m(\theta))
\end{equation}
where $N$ is the number of data points, $\Sigma$ is the covariance matrix, $d$ is the observed data, $m(\theta)$ is the model given the parameters $\theta$.

The covariance matrix $\Sigma$ is calculated from the uncertainties of the data, the correlated noise between pixels and the local deviations from the mean of the residuals of the fit. We construct a covariance matrix adapted from \citealt{czekalaConstructingFlexibleLikelihood2015} that accounts for correlated noise and provides a robust framework to manage local residual features. In summary, the covariance matrix is calculated as $\Sigma = \Sigma_{\rm obs} + \Sigma_{\rm corr} + \Sigma_{\rm out}$. The individual components of the covariance matrix are defined as:

\begin{align}
    \Sigma_{\rm obs} &= \delta_{ij} \sigma^2_{\rm obs} \times 10^{2b} \\
    \Sigma_{\rm corr} &= \sigma_{\text{eff}}^2 \left(1 + \frac{\sqrt{3}|x_i-x_j|}{l_{\rm G}}\right) \exp\left(-\frac{\sqrt{3}|x_i-x_j|}{l_{\rm G}}\right) \\
    \Sigma_{\rm out} &= \sum_{k} a_{k}^2 \exp\left(-\frac{(x_i-x_k)\cdot (x_j-x_k)}{2l_k^2}\right)
\end{align}
The first term is the diagonal covariance of the observational uncertainties, scaled by a factor $10^{2b}$ to account for underestimated uncertainties. A separate factor $b_\text{grating}$ is applied to each grating to account for the different noise levels at different wavelengths. 

The second term accounts for correlated noise in the data which may be present due to instrumental effects (e.g. fringing or interpolation errors) or due to residual features from a mismatch between the model and the data. We model this with a Mat\'ern kernel of order 3/2 \citep{maternSpatialVariation1986} with amplitude $\sigma_{\text{eff}} = 10^b \times \text{median}(\sigma_i)$ and global length scale $l_{\rm G}$ in units of velocity.

The third term accounts for outliers in the data that may introduce local correlated noise \citep{czekalaConstructingFlexibleLikelihood2015}. We model this using a Gaussian kernel with amplitude $a_k$ and fixed length scale $l_k$. Outlier locations and amplitudes are identified from the fit residuals at each model evaluation. Outliers are flagged as pixels with a chi-square per pixel $\chi_i = (d_i - m_i)\Sigma_{\rm obs+corr, ij}^{-1}(d_i - m_i)$ exceeding $6\sigma$. To preserve the definition of outliers, we limit the number density of local kernels to a maximum of five per 40~nm. Based on initial tests, we fix the kernel length scale to $l_k = 30$~km~s$^{-1}$, which effectively captures the extent of local outliers. The residuals are overplotted with the square root of the diagonal of the fitted covariance matrix (see Appendix~\ref{sec:best_fit_model_spectra}).

\section{Results}\label{sec:results}
Our atmospheric retrievals of TWA 27A and TWA 28 provide robust constraints on composition and temperature structure. \Cref{fig:full_spec} shows the best-fit models across the full wavelength range, with detailed spectral segments and opacity contributions in \Cref{fig:spec_opacities_1,fig:spec_opacities_2}. Temperature profiles appear in \Cref{fig:PTs}, molecular detections via cross-correlation functions in \Cref{fig:ccf}, and all parameters with uncertainties in \Cref{tab:free_params_comparison_0}.

\subsection{Chemical composition}\label{sec:results_chemical_composition}
Our retrievals constrain more than twenty molecular and atomic species. We conduct a comprehensive analysis identifying the dominant molecular and atomic species in the different regions of the spectra (see \Cref{fig:spec_opacities_1,fig:spec_opacities_2}). The vertical composition profiles for the major atmospheric species are shown in \Cref{fig:VMRs}.

At the reddest part of the spectra, the fundamental CO band enables the detection of the minor isotopologues of CO: \thirteenCO{}, \CeighteenO{} and tentative evidence of \CseventeenO{}. At 4.0--5.2~\micron{}, \twelveCO{}, \water{}, \COtwo{}, and SiO dominate the opacity (\Cref{fig:ccf}).

Water dominates the absorption features between 2.5 and 4.0~\micron{}. The isotopologue \eighteenOwater{} is constrained across the entire wavelength range, with distinct lines identified from 2.5~\micron{} towards redder wavelengths. Lines from OH, HF, and HCl are detected in this region too. Methane (\methane) is tentatively detected, with broad absorption near 3.3~\micron{}. This detection remains tentative due to residuals in this region (see \Cref{fig:full_spec}) and a highly non-equilibrium abundance for \methane{} (see \Cref{tab:free_params_comparison_1}).

In the 2.0--2.5~\micron{} range, we observe the first overtone of \twelveCO{} and weak features from \thirteenCO{}. This region also shows lines from Na, Ca, Ti, OH, and HF. The water absorption peak at 1.65~\micron{}, a proposed gravity-sensitive feature, is clearly visible.

Between 1.4 and 2.0~\micron{}, absorption from K, Al, and Ca is present, with a strong Ca signature at 1.96~\micron{}. From 0.97 to 1.4~\micron{}, we identify hydrides, oxides, and atomic features. FeH, TiO, and VO show strong absorption, along with Na and K doublets. We also constrain CrH, NaH, and AlH, and find evidence for AlO, which we retrieve as a free parameter due to its absence in our chemical equilibrium model. At the bluest region of the spectrum, the residuals exhibit significant structure, suggesting missing opacity sources or other unaccounted for effects.

We identify a continuum contribution from H$^-$ bound-free opacity at wavelengths shorter than 1.64~\micron{} \citep{grayObservationAnalysisStellar2022}. The retrieved H$^-$ abundance, $\log{\text{H}^{-}} \approx -8.84 \pm 0.03$, is consistent across both objects and matches chemical equilibrium predictions at relevant temperatures. This wavelength range probes deeper atmospheric layers. Our use of a constant-with-altitude abundance profile limits detailed interpretation, but the retrieved H$^-$ abundance should represent photospheric conditions.

\subsection{Temperature profile}\label{sec:results_temperature_profile}
\begin{figure}
    \centering
    \includegraphics[width=\columnwidth]{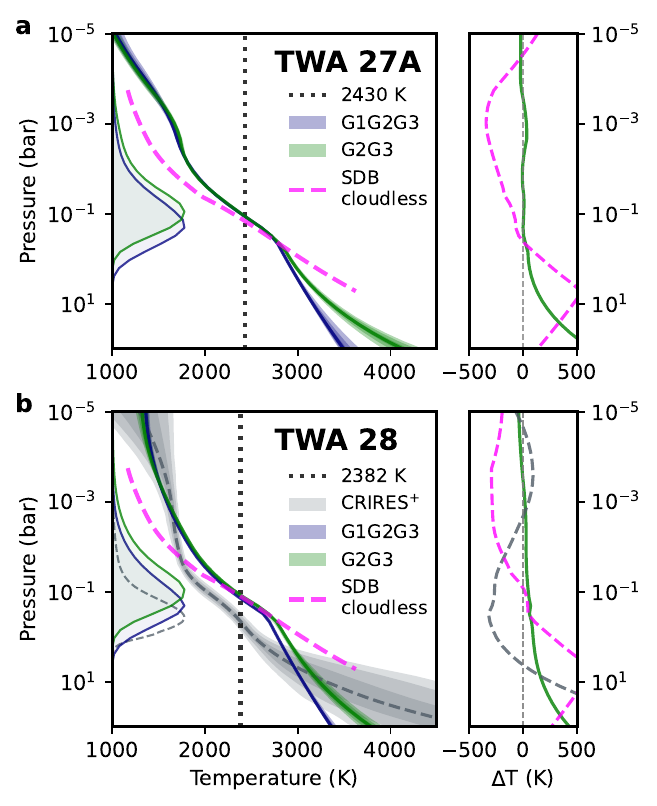}
    \caption{Atmospheric temperature profiles for TWA 27A (\textbf{a}) and TWA 28 (\textbf{b}). Left panels: Retrieved profiles from JWST/NIRSpec observations (G140H+G235H+G395H, blue lines) and excluding G140H (green lines), with 1-, 2-, and 3-$\sigma$ confidence intervals (shaded regions). CRIRES+ profile for TWA 28 \citep{gonzalezpicosESOSupJupSurvey2024} and a cloudless Sonora Diamondback model at $T_{\mathrm{eff}}=2400$ K, $\log{g}=4.0$ \citep{morleySonoraSubstellarAtmosphere2024} are shown for comparison. Right panels: Residuals between median G140H+G235H+G395H profiles and other datasets.}
    \label{fig:PTs}
\end{figure}

We fit seven temperature gradients at pressure levels determined by three free parameters to recover the temperature profile. \Cref{fig:PTs} shows the resulting profiles with confidence intervals. TWA 28 is marginally cooler than TWA 27A, consistent with previous effective temperature estimates \citep{cooperUltracoolSpectroscopicOutliers2024}. The tight constraints on the shape of the temperature profile result from the broad wavelength coverage and high S/N of the data. We compare with previous high-resolution $K$-band analysis of TWA 28 in \Cref{sec:comparison_with_cries}.

\subsection{Detection of excess infrared emission}\label{sec:excess_infrared_emission}
Both objects show excess infrared emission beyond 2.5~\micron{}. We model this as an optically thick ring with single-temperature blackbody emission, retrieving the temperature and radius of the emitting region. \Cref{fig:full_spec} (upper right inset) illustrates this continuum contribution.
\begin{align*}
    \text{TWA 28}: T^{\rm bb}_{\mathrm{eff}} = 653\pm 2\,{\mathrm{K}},\,R^{\rm bb} = 13.9 \pm 0.08 \,{\mathrm{R}}_{\!\mathrm{Jup}}\\
    \text{TWA 27A}: T^{\rm bb}_{\mathrm{eff}} = 643\pm 4\,{\mathrm{K}},\,R^{\rm bb} = 11.9 \pm 0.15 \,{\mathrm{R}}_{\!\mathrm{Jup}}
\end{align*}

The retrieved values support the presence of warm dust in the inner disk, consistent with previous studies that identified infrared excess in these sources \citep{boucherBANYANVIIINew2016,venutiXshooterSpectroscopyYoung2019,manjavacasMediumresolution09753Mm2024}, with \citealt{boucherBANYANVIIINew2016} estimating disc temperatures $>300$~K, consistent with our values (see \Cref{sec:disc_properties}).

\begin{figure}
    \centering
    \includegraphics[width=\columnwidth]{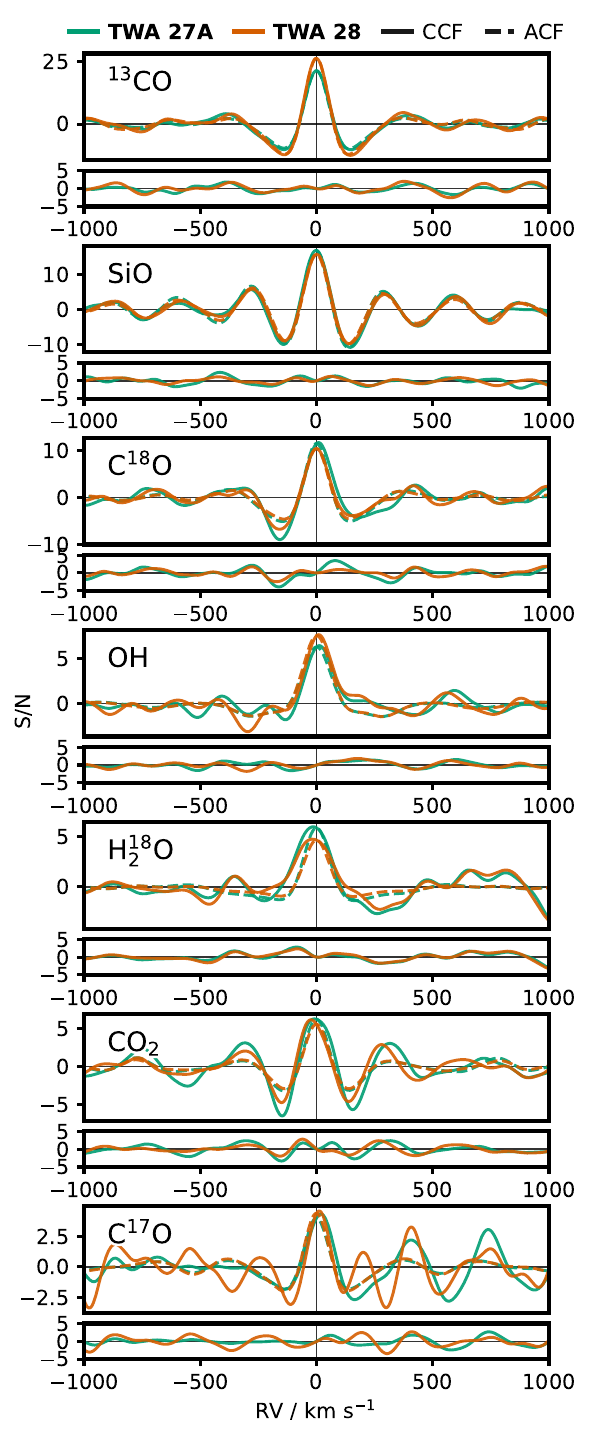}
    \caption{Molecular detections via cross-correlation analysis. CCF of selected molecules for TWA 27A and TWA 28, computed between model residuals (with and without each species) and template spectra over radial velocities up to 2000~km~s$^{-1}$. S/N ratios are calculated by normalising the CCF peak to the standard deviation of the CCF-ACF difference (secondary panels). ACF is the autocorrelation function of each molecule.}
    \label{fig:ccf}
\end{figure}

\subsection{Line emission from the disc}\label{sec:disc_line_emission}
Strong residuals in the CO fundamental band (\cref{fig:ring_emission}, middle panel) from preliminary fits with atmospheric models suggest the presence of line emission in both systems. Atmospheric models alone cannot reproduce the observed CO line depths. Adding a slab component improves the fit to the data and leads to reasonable abundances of atmospheric \twelveCO. We constrain the column densities, excitation temperatures, sizes of emitting regions and the radial velocity of the slab models (see \Cref{fig:corner_plot_TWA27A_TWA28}). We present these values in the context of disc observations and theoretical models in \Cref{sec:slab_model}.

\section{Discussion}\label{sec:discussion}

\subsection{Deviations from solar-like chemical equilibrium}\label{sec:deviations_from_chemical_equilibrium}
\begin{figure}
    \centering
    \includegraphics[width=\columnwidth]{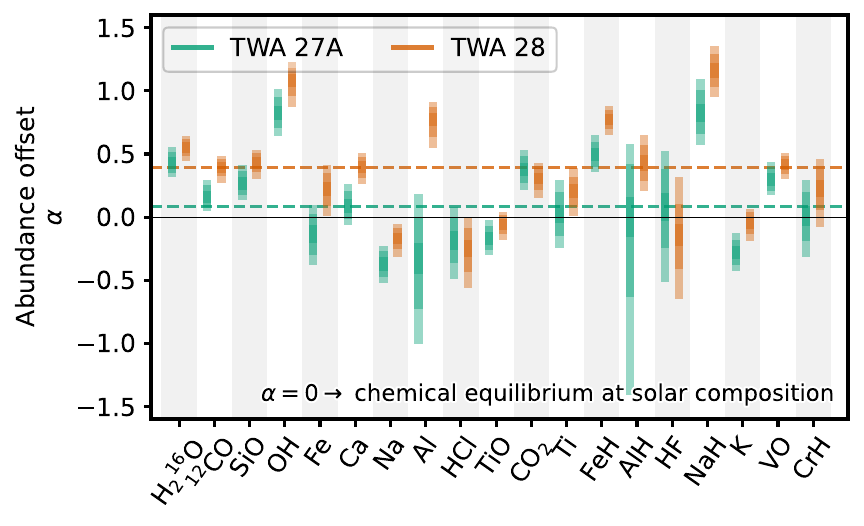}
    \caption{Chemical abundance offsets from solar-composition chemical equilibrium. The 1-, 2-, and 3-$\sigma$ confidence intervals of $\alpha$ as defined in \Cref{eq:alpha} are shown for each of the retrieved species using chemical equilibrium models. The median value for each object is plotted as a horizontal dashed line.}
    \label{fig:alpha}
\end{figure}

We estimate departures from chemical equilibrium for each species using the $\alpha$-parameter (see \Cref{fig:alpha}). Most species show positive $\alpha$ values, indicating super-solar metallicity or slight discrepancies with equilibrium chemistry. TWA 27A and TWA 28 exhibit consistent $\alpha$ values, reflecting their chemical similarity—unsurprising given their matched spectral types, ages, and distances.

Notably high values of $\alpha$, such as those retrieved for OH, suggest significant departures from chemical equilibrium. The formation of OH through thermal dissociation of H$_2$O is known to be efficient at temperatures exceeding 2000 K. A value of $\alpha_{\text{OH}} = 1.0$ indicates that the retrieved OH abundance is 10 times higher than that predicted by chemical equilibrium at solar-metallicity.

\begin{figure*}[h!]
    \centering
    \includegraphics[width=\textwidth]{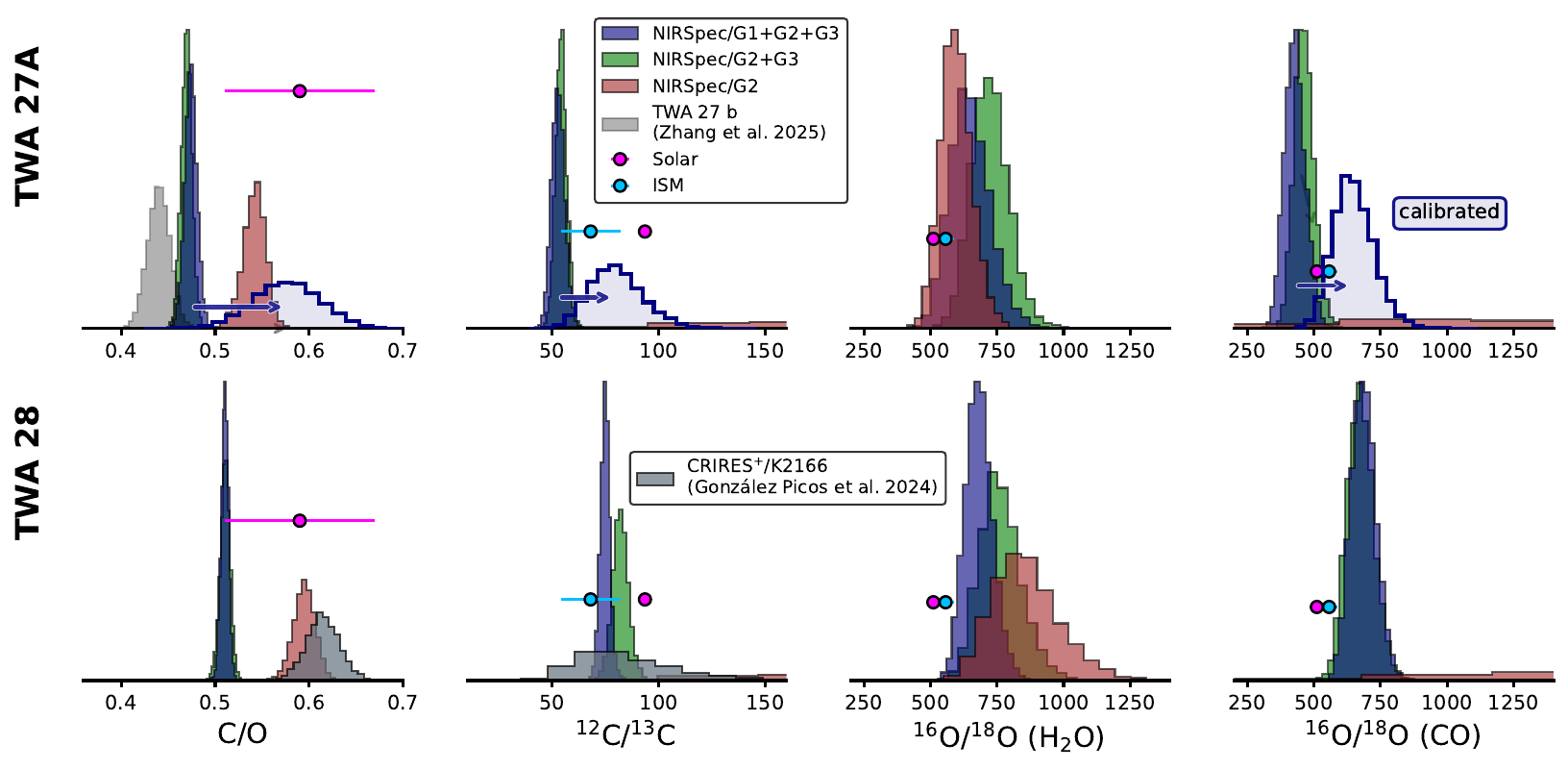}
    \caption{Posterior distributions of atmospheric parameters retrieved from the JWST/NIRSpec observations of TWA 27A (top row) and TWA 28 (bottom row). From left to right, the panels show the carbon-to-oxygen ratio, carbon isotope ratio, oxygen isotope ratio from \water{} and from CO. The vertical axis of each panel represents the posterior probability density (not shown). The oxygen-isotope homogeneity calibration is applied to TWA 27A, with the resulting posteriors indicated with an arrow. Reference values for the ISM, the solar value and literature values are shown where available. Results from JWST/NIRSpec for the planetary-mass companion of TWA 27A  \citep{zhangELementalAbundancesPlanets2025} and for TWA 28 from HRS \CRIRES{} observations \citep{gonzalezpicosESOSupJupSurvey2024} are shown in gray.}
    \label{fig:carbon_oxygen_isotope_ratios_calibrated}
\end{figure*}

From the abundances of carbon- and oxygen-bearing species, we infer C/O ratios for each dataset (\Cref{fig:carbon_oxygen_isotope_ratios_calibrated}, left panel). Including the reddest grating (G395H) produces lower C/O ratios than using only the intermediate grating (G235H). This difference stems from atmospheric \twelveCO{} absorption competing with warm disc emission in the CO fundamental band.

We consider the C/O values derived from the G235H grating to be more representative of the intrinsic atmospheric composition, as this region is unaffected by the CO line emission from the disc at a significant level (see inset of \Cref{fig:full_spec}). The resulting values are:
\begin{equation}
    \text{C/O}_{\text{TWA 27A}} = 0.54 \pm 0.02 \quad \text{and} \quad \text{C/O}_{\text{TWA 28}} = 0.59 \pm 0.02.
\end{equation}
These are comparable to the solar value of $0.59 \pm 0.08$ \citep{asplundChemicalMakeupSun2021}. A recent analysis of the planetary-mass companion in the TWA 27 system using JWST/NIRSpec \citep{zhangELementalAbundancesPlanets2025} found that a partial cloud deck atmosphere model best fits the spectrum of TWA 27b, with a retrieved sub-solar C/O ratio of $0.440 \pm 0.012$. This value differs from our measurement for the host brown dwarf by approximately $3\sigma$, suggesting a potential slight chemical difference between the host brown dwarf and its companion. However, we note the difference is small and the uncertainties might be underestimated for one or both objects.

Although retrievals including the G395H grating consistently favour lower C/O values, we caution that these are likely affected by CO slab emission. Consequently, they may underestimate the true atmospheric C/O ratio. High resolution spectroscopy might be able to provide a more robust measurement of the C/O ratio \citep[e.g.][]{costesFreshViewHot2024,zhangAtmosphericCharacterizationSuperJupiter2024}.

\begin{figure*}[h!]
    \centering
    \includegraphics[width=\textwidth]{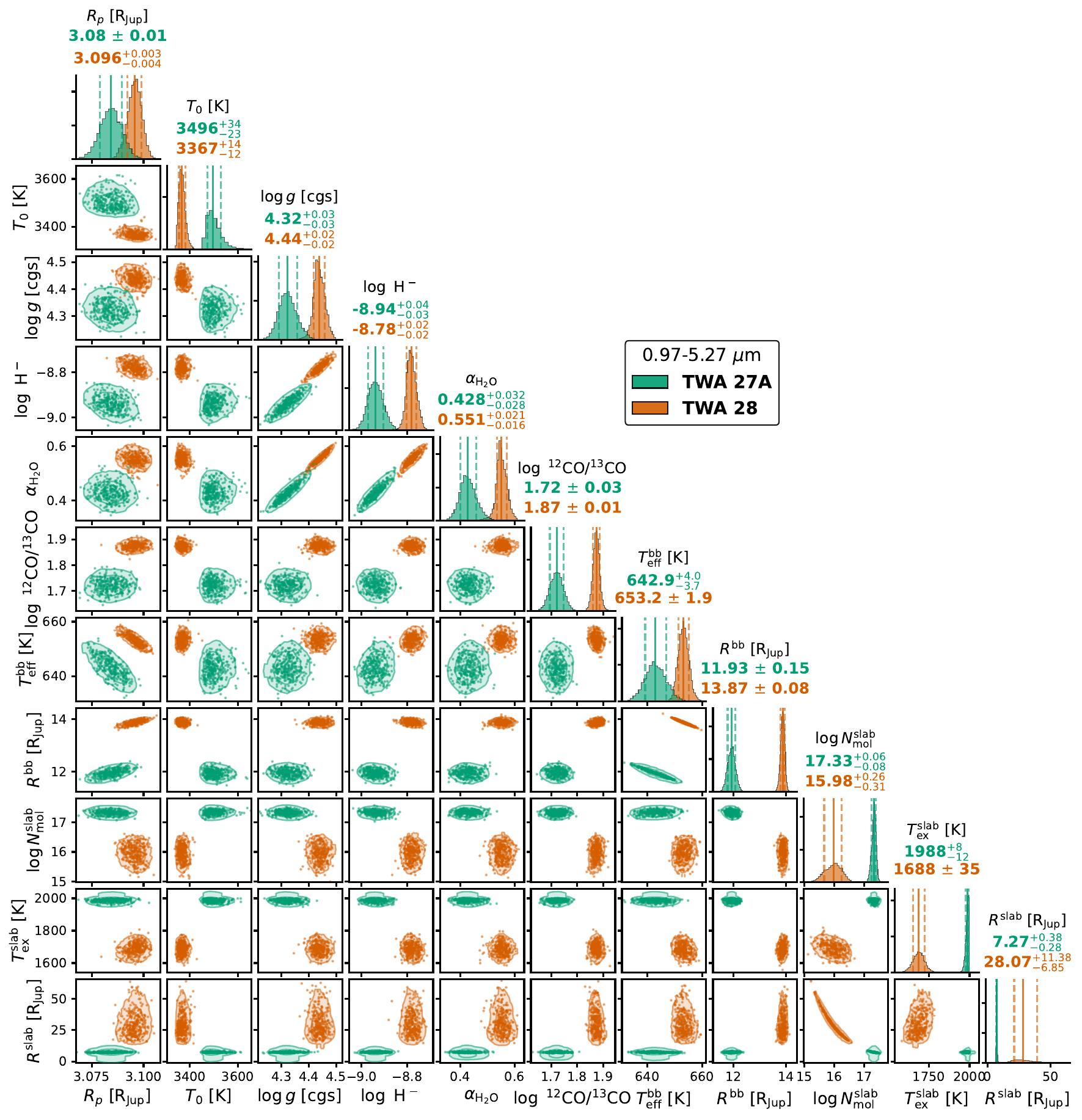}
    \caption{Corner plot of the posterior distributions of selected parameters for TWA 27A and TWA 28. The parameters are the radius, surface temperature, surface gravity, logarithm of the volume mixing ratio of the H$^{-}$ bound-free opacity, the $\alpha$-parameter of water, the carbon isotopologue ratio, the effective temperature and size of the blackbody, and the logarithm of the column density, the excitation temperature and the size of the slab model. The titles indicate the median and the 16th and 84th percentiles of the posterior distributions for each target retrieved from the fit to the entire wavelength range.}
    \label{fig:corner_plot_TWA27A_TWA28}
\end{figure*}

\subsection{Isotope ratios}\label{sec:isotope_ratios}
We report detections of several minor isotopologues of CO ($^{13}$CO, C$^{18}$O, C$^{17}$O) and the secondary isotopologue of water, H$_2^{18}$O (see \Cref{fig:ccf}). The fundamental CO band at 4.3--5.1~\micron{}, covered by the reddest NIRSpec grating (G395H), enables robust constraints on the CO isotopologues. In contrast, retrievals using only the G235H grating yield no significant detections of \thirteenCO{} and provide no constraints on \CeighteenO{} or \CseventeenO{}. Notably, the retrieved isotope ratios show minimal dependence on surface gravity and metallicity (see \Cref{fig:corner_plot_TWA27A_TWA28}).

In \Cref{tab:isotope_ratios}, we present the retrieved isotopologue ratios for each object. The inferred \Cratio{} from CO is lower for TWA~27A ($53\pm3$) compared to TWA~28 ($75\pm2$). For the oxygen isotope ratios (\Oratio{}), we find consistent values between \water{} and CO for TWA~28. However, TWA~27A shows a notable discrepancy, with the \Oratio{} from \water{} being significantly higher than that from CO. This suggests that the \twelveCO{} abundance in TWA~27A is underestimated, leading to systematically lower carbon and oxygen isotope ratios when derived from CO alone.

\begin{table}
    \centering
    \caption{Isotope ratios derived from JWST/NIRSpec observations}
    \label{tab:isotope_ratios}
    \begin{tabular}{lcc}
    \hline\hline
    \rule{0pt}{3ex}Target & TWA 27A & TWA 28 \\[1ex]
    \hline
    \rule{0pt}{3ex}
    \rule{0pt}{1ex}$^{12}$CO/$^{13}$CO & $53_{-3}^{+3}$ & $75_{-2}^{+2}$ \\[1ex]
    \rule{0pt}{1ex}$^{12}$CO/$^{13}$CO (calibrated) & $79_{-11}^{+14}$ & --- \\[1ex]
    \rule{0pt}{1ex}$^{12}$CO/C$^{18}$O & $428_{-37}^{+42}$ & $685_{-46}^{+49}$ \\[1ex]
    \rule{0pt}{1ex}H$_2^{16}$O/H$_2^{18}$O & $645_{-70}^{+80}$ & $681_{-50}^{+53}$ \\[1ex]
    \hline
    \end{tabular}
    \tablefoot{The uncertainties represent the 16th and 84th percentiles of the posterior distributions. The calibrated $^{12}$CO/$^{13}$CO ratios account for oxygen isotope homogeneity using the gamma factor $\gamma = ({\rm H_2^{16}O/H_2^{18}O}) \times ({\rm ^{12}CO/C^{18}O})^{-1}$ to ensure consistent oxygen isotope ratios between CO and H$_2$O molecules. For TWA 28, $\gamma \approx 1$ within uncertainties, so no calibration is applied.}
\end{table}

\begin{figure}
    \centering
    \includegraphics[width=\columnwidth]{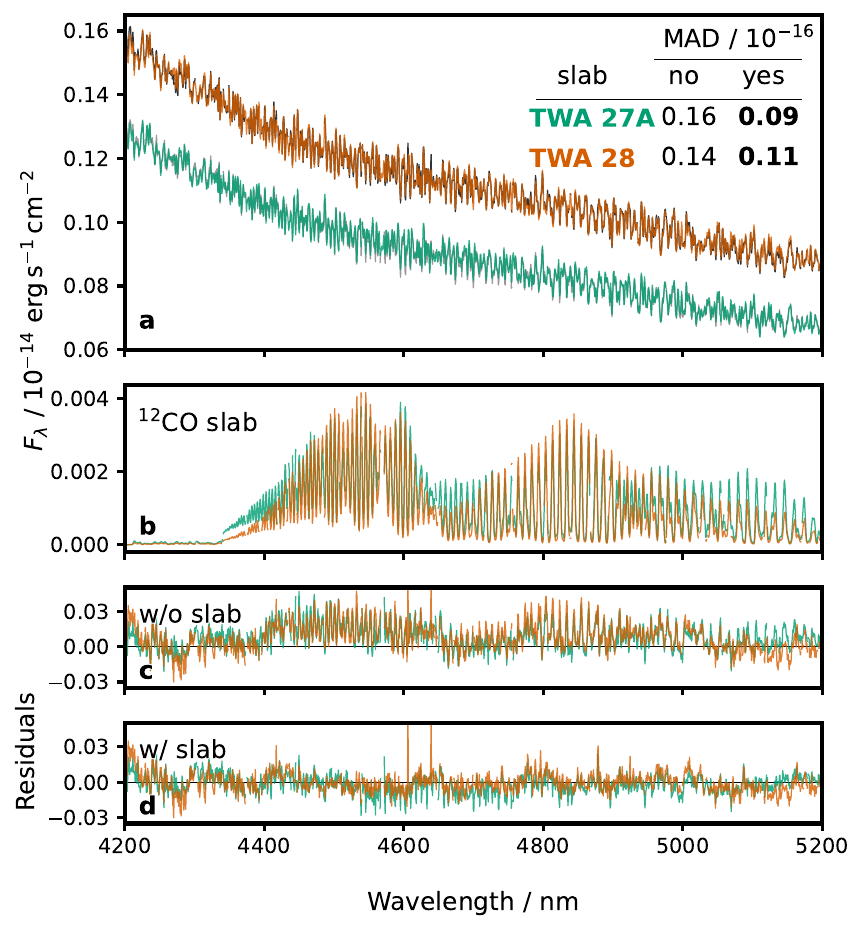}
    \caption{Fundamental CO band with disc slab emission. \textbf{a}. Best--fit spectra of TWA~27A and TWA~28 the CO band. \textbf{b}. Retrieved slab emission. \textbf{c, d}. Residuals of the fit excluding (\textbf{c}) and including (\textbf{d}) the slab model. The legend lists the median absolute deviation (MAD); lower MADs suggest a superior fit.}
    \label{fig:ring_emission}
\end{figure}

This interpretation is supported by prominent residuals in the CO band region of the TWA~27A spectrum, which are absent in the TWA~28 data (see \Cref{fig:ring_emission}cd). We estimate a calibration factor to the \twelveCO{} abundance in TWA~27A by assuming that the \Oratio{} values from \water{} and CO should be homogeneous, as generally expected and supported by the consistency seen in TWA~28. Since the \Oratio{} measured from \water{} is not affected by the CO slab emission, it provides a reliable proxy for the true isotopic composition. The retrieved \Oratio{} values for both objects agree within uncertainties and lie close to or slightly above the interstellar medium value ($557 \pm 30$; \citealt{wilsonIsotopesInterstellarMedium1999}) and the solar value ($511 \pm 10$; \citealt{ayresSUNLIGHTEREARTH2013}).

We find that the \twelveCO{} abundance for TWA~27A is underestimated by a factor of $\approx 1.51$. Propagating this correction to the derived C/O and carbon isotope ratios results in:
\begin{align*}
    \text{C/O}_{\text{TWA 27A}} & = 0.58 \pm 0.04 \\
    ^{12}\text{C}/^{13}\text{C}_{\text{TWA 27A}} & = 79^{+14}_{-11}
\end{align*}
This calibration brings the C/O ratio into agreement with the value derived from the G235H grating and aligns the carbon isotope ratio with TWA~28 within $2\sigma$. The increased uncertainty on the carbon isotope ratio is due to the propagated errors on the calibration factor that incorporates the spread in the \Oratio{} from \water{} and CO. Although this correction is approximate, it demonstrates the value of using multiple isotopologues to constrain chemical abundances. Additionally, we find tentative evidence for \CseventeenO{} in our spectra. Cross-correlation analysis indicates a signal-to-noise ratio of approximately 3 for this detection (see \Cref{fig:ccf}); however, the derived \ratio{O}{16}{17} is poorly constrained and we opt to report a lower limit of $\approx$1000 for both objects.

We investigate the presence of spectral features from TiO isotopologues in the bluest region of the spectrum (0.97--1.63~\micron). We perform two sets of retrievals: one including a line list with all stable isotopologues of TiO assuming terrestrial isotope ratios \citep{loddersAbundancesElementsSolar2009,mckemmishHybridApproachGenerating2024}, and another using only the $^{48}$TiO line list \citep{mckemmishHybridApproachGenerating2024}. To ensure a fair comparison, we fix the temperature profile to that obtained from the fit to the entire wavelength range (see \Cref{fig:PTs}). We find slight differences in the resulting parameters between the two models, but they are generally consistent within 1$\sigma$ uncertainties. The median absolute deviations of both fits do not change significantly, suggesting that the addition of secondary isotopologues does not clearly improve the fit, at least not at the fixed contribution set by terrestrial values. The variations from TiO isotopologues appear to be below the current noise threshold imposed by the mismatch between data and model. Future studies searching for Ti isotopes in TiO might benefit from improved fits in this region through the inclusion of additional oxides and hydrides. Alternatively, observations at other wavelength regions, such as the optical range covering the 0.75~\micron{} band of TiO, and at higher resolving power, might be more amenable to the measurement of these elusive isotopologues \citep{pavlenkoAnalysisTiOIsotopologues2020,serindagTiOEmissionPresent2021}.

\subsection{Disc properties}\label{sec:disc_properties}
The wide wavelength coverage and simultaneous fit of the atmosphere and disc enable the characterisation of the excess continuum and line emission properties. 

\subsubsection{Continuum emission}
We find that the excess continuum emission of both objects is well described by a ring-like structure emitting as a single-temperature blackbody at approximately 650~K for both objects (see values in \Cref{tab:free_params_comparison_0}). The size of the disc appears slightly larger for TWA 27A compared to TWA 28, but this may result from the inclination of the system rather than the actual size of the disc. The retrieved sizes of these discs are small compared to other discs around young stars, which typically span a radial range of several dozens of au \citep[e.g.][]{huangCODustProperties2018}. This result is consistent with ALMA observations of the TWA 27 system that report a compact disc \citep{ricciALMAObservationsYoung2017}. The inferred blackbody temperature of both discs is remarkably similar, identical within 1$\sigma$ uncertainties. These values agree with previous near-infrared observations of the same systems by \citet{boucherBANYANVIIINew2016} that report disc temperatures of approximately 250--850~K. 

We estimate the disc cavity size using the Stefan-Boltzmann relation following Eq.~1 from \citet{cugnoMidInfraredSpectrumDisk2024}. Using effective temperatures from \citet{cooperUltracoolSpectroscopicOutliers2024} and our retrieved radii ($R = 3.08 \pm 0.01$~\Rjup{} and $3.10 \pm 0.01$~\Rjup{}, respectively), we calculate bolometric luminosities of $\log(L_{\mathrm{bol}}/L_{\odot}) = -2.48 \pm 0.02$ (TWA 27A) and $-2.54 \pm 0.03$ (TWA 28). The cavity radii, estimated as $R_{\mathrm{cavity}} = \sqrt{L_{\mathrm{bol}} / (16\pi \sigma T_{\mathrm{bb}}^4)}$ where $T_{\mathrm{bb}}$ is the effective temperature of the blackbody disc from our retrievals (see \Cref{sec:excess_infrared_emission}), are $22.50 \pm 0.64$~\Rjup{} and $20.49 \pm 0.74$~\Rjup{} for TWA 27A and TWA 28, respectively. These estimates suggest the presence of substantial inner disc cavities, consistent with observations of other young substellar objects \citep{cugnoMidInfraredSpectrumDisk2024}. We note that these are rough estimates that assume simplified disc geometry and neglect potential heating effects. Notably, discs around similar analogue substellar companions show comparable disc temperatures and sizes, namely GQ Lup B ($T_{\mathrm{disc}} = 575$--$630$~K, $R_{\mathrm{disc}} = 23$--$28$~R$_{\mathrm{J}}$; \citealt{cugnoMidInfraredSpectrumDisk2024}) and YSES 1b ($T_{\mathrm{disc}} = 320$--$520$~K, $R_{\mathrm{disc}} = 5.6$--$18$~R$_{\mathrm{J}}$; \citealt{hochSilicateCloudsCircumplanetary2025}).

We extend our best-fit model to longer wavelengths to assess the validity of the blackbody approximation in regions where the disc contribution becomes more prominent. In \Cref{fig:spec_spitzer} we show that our models are in good agreement with Spitzer/IRS observations \citep{riazNewBrownDwarf2008} up to 11 and 12 \micron{} for TWA 27A and TWA 28, respectively. This suggests the single-temperature blackbody correctly describes the dominant continuum emission from surrounding dust at the wavelengths considered in the present work, but additional emission from colder dust may be necessary to match observations in the mid-infrared.

\subsubsection{Line emission}
We constrain the excitation temperature of both objects, finding a value of $T_{\mathrm{ex}}(\mathrm{TWA\ 28}) = 1688 \pm 36$~K and a lower limit of $T_{\mathrm{ex}}(\mathrm{TWA\ 27A}) > 2000$~K (the edge of our prior). These temperatures are significantly higher than the continuum temperature of 650~K, suggesting that the CO gas emission originates in a different part of the disc. Elevated excitation temperatures have been observed in other systems \citep[e.g.][]{temminkMINDSDRTau2024} and are often attributed to emission in non-LTE conditions, namely UV pumping \citep{krotkovUltravioletPumpingMolecular1980}. For these brown dwarfs (Teff ~2400K), the UV flux likely originates from nearby stars in the TWA association rather than from the brown dwarfs themselves, which are too cool to produce significant UV emission. Notably, \citet{bastSinglePeakedCO2011} reported observations of UV-excited CO vibrational emission in several protoplanetary discs at temperatures around 1700 K. The slab modelling used in this work assumes LTE conditions \citep{munoz-romeroRetrievalThermallyResolvedWater2024}, which may lead to limited physical interpretation of slab parameters. The retrieved column densities (see \Cref{tab:free_params_comparison_0}) are in the range of $10^{16}$--$10^{18}$~cm$^{-2}$ for both objects, which is the order of magnitude range observed in other substellar discs \citep[e.g.][]{pascucciATOMICMOLECULARCONTENT2013}. However, we refrain from further interpretation due to the strong correlation with the size of the emitting region (see \Cref{fig:corner_plot_TWA27A_TWA28}). JWST/MIRI detected \COtwo{} emission around 15~\micron{} in TWA 27A's inner disc \citep{arabhaviMINDSVeryLowmass2025}. That study found no CO emission—likely because photospheric absorption and disc emission overlap, complicating line identification as also underscored in the present work (see \Cref{sec:isotope_ratios}).

In addition, we detect a velocity shift of the slab line emission with respect to the photospheric lines of $8 \pm 2$ and $4 \pm 1$~km~s$^{-1}$ for TWA 27A and TWA 28, respectively. This velocity shift is consistent with the range of velocities determined by the spatially integrated CO emission observed by ALMA \citep{ricciALMAObservationsYoung2017}. At the resolving power of JWST/NIRSpec, the atmospheric and disc CO emission is unresolved, but velocity shifts of a few km~s$^{-1}$ are within reach of current high-resolution ground-based spectroscopy \citep[e.g.][]{grantFullMbandHigh2024}.

\subsection{Missing opacity sources}\label{sec:missing_opacity_sources}
The best-fit model spectrum reproduces most observed spectral features, while the residuals show an overall scatter at a level of 5\% (see residuals panel in \Cref{fig:full_spec}). However, significant residuals remain at certain wavelengths, particularly in the blue part of the spectrum and around 2.9-3.5 \micron{}, showing two apparently broad, unidentified features. These structured residuals suggest missing molecules in our chemical inventory or inaccurate line lists. We discuss our attempts to account for these features:

-- 0.97-1.30 \micron{}: This region is dominated by H$_2$O, hydrides (FeH, HF, CrH, NaH), oxides (TiO, VO), and atomic species (Na, K, Fe). We tested additional atoms (Mg, Al, Si, Mn, Cr, V, Cs, Li) but found no constraints on their abundances. Some residual features show molecular bandhead structure, but none of the known opacity of SiH, NH, SH, CH, MgO, or ZrO matched the observed features.

-- 1.59-1.71 \micron{}: The dominant opacity source is H$_2$O, followed by $^{12}$CO, OH, and a weak contribution from FeH. Several absorption lines in the data remain unaccounted for, with residuals similar to those seen in high-resolution spectra of a nearby M4 dwarf star \citep{jahandarComprehensiveHighresolutionChemical2023}.

-- 2.9-3.6 \micron{}: This region is primarily dominated by H$_2$O, with additional contributions from OH, HCl, and HF. We identify two prominent dips (D1: 2960-3140 nm and D2: 3360-3520 nm) where the data points fall significantly below the model. We consider these features to be real rather than instrumental effects because both gratings G235H and G395H cover the D1 feature and show remarkable agreement. We rule out fringing as a possible instrumental systematic to explain the double-peak structure, as analysis of different dithering positions and extraction apertures shows no evidence for fringing effects at the level of the residuals \citep{dumontWIggleCorrectorKit2025}.

We tested several carbon- and nitrogen-bearing molecules that could contribute to these features, including CH$_4$, NH$_3$, HCN, C$_2$H$_2$, CH, NH, SH, SiH, H$_2$S, MgO, AlO, and SiO. While CH$_4$ and C$_2$H$_2$ improved the fit, they required abundances 2-3 orders of magnitude higher than predicted by chemical equilibrium. C$_2$H$_2$ partially matched the D1 feature; however, the presence of this molecule above 2000 K is highly disfavoured by thermochemical models \citep{kitzmannFastchemCondEquilibrium2023}. CH$_4$ improved the overall fit but did not reproduce either dip, instead adding uniform opacity. In our final retrievals, we included CH$_4$ as a tentative opacity source but note that it is also highly disfavoured by chemical equilibrium. The detection of C$_2$H$_2$ and HCN in \citep{matthewsHCNC2H2Atmosphere2025} suggests that these molecules might be present in the atmosphere of substellar objects despite the negligible abundances predicted by chemical equilibrium, but may be explained by ionisation effects \citep{hellingLightningChargeProcesses2019}.

\subsection{Surface gravity and absolute abundances}\label{sec:surface_gravity_absolute_abundances}
\begin{figure}[h]
    \centering
    \includegraphics[width=\columnwidth]{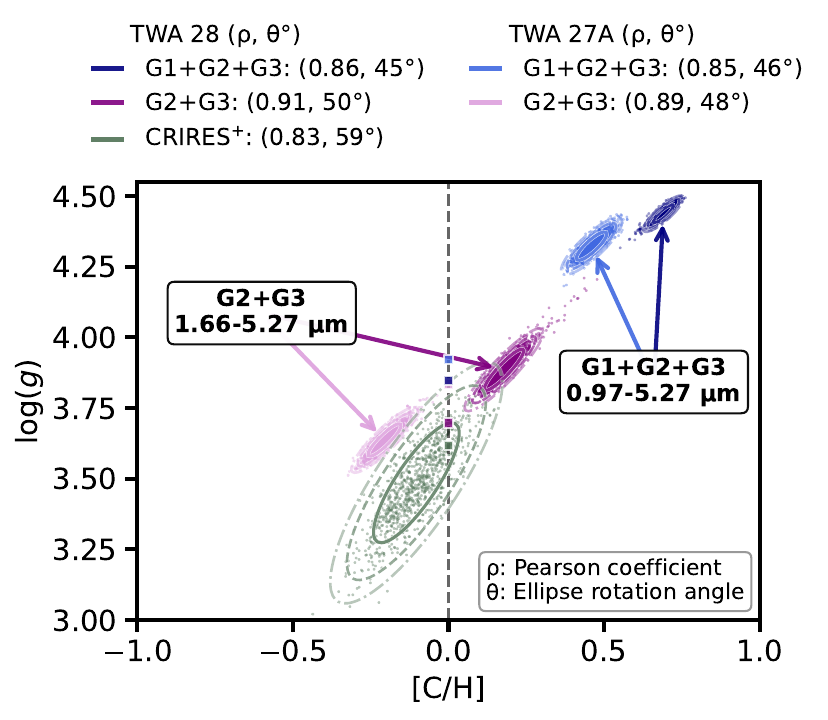}
    \caption{Surface gravity-metallicity degeneracy analysis. Posterior distributions of the surface gravity and the inferred metallicity. The \CRIRES{} results from GP24 are included for reference. Ellipses show the 68\% and 95\% and 99\% confidence regions. The legend lists the Pearson correlation coefficient ($\rho$) and the rotation angle ($\theta$) of the ellipse.}
    \label{fig:logg_metallicity_correlation}
\end{figure}

Our atmospheric retrieval simultaneously fits radius, surface gravity, and blackbody disc emission parameters. The retrieved surface gravities exhibit strong dependence on the selected wavelength range (\cref{fig:logg_metallicity_correlation}) and show significant correlation with the abundances of the main opacity sources or metallicity (\cref{fig:logg_metallicity_correlation}), similar to previous findings from high-resolution spectroscopy of TWA 28 \citep{gonzalezpicosESOSupJupSurvey2024}. We note that the metallicity values shown in \Cref{fig:logg_metallicity_correlation} consider only carbon-bearing species, as we fit individual alpha parameters for each species rather than a global metallicity. This approach allows us to capture species-specific deviations from chemical equilibrium while maintaining the flexibility to model complex atmospheric chemistry.

Notably, including the bluest spectral region (G140H, 0.97--1.89~\micron{}) drives both surface gravity and metallicity to higher values compared to fits using only the redder spectrum (1.66--5.27~\micron{}). Although several alkali lines in the G140H region are proposed as gravity-sensitive indicators, the poor model fit in this spectral region may introduce systematic biases in the retrieved surface gravity values.

To assess the consistency of solutions across different wavelength subsets, we examine the linear correlation between surface gravity and metallicity for various data combinations (\cref{fig:logg_metallicity_correlation}). The corresponding values of surface gravities at [C/H]=0 as extrapolated from the correlation line remain broadly consistent: $\logg{} \approx 4.10$--$4.17$ and $\logg{} \approx 4.03$--$4.10$ for TWA 27A and TWA 28, respectively. This suggests compatible solutions despite the presence of strong parameter degeneracies and likely underestimated uncertainties.

\subsection{Comparison with \CRIRES observations of TWA 28}\label{sec:comparison_with_cries}
High-resolution \textit{K}-band spectroscopy of TWA 28 was obtained with the upgraded CRyogenic InfraRed Echelle Spectrometer at the Very Large Telescope (\citealt{dornCRIRESSkyESO2023}; VLT/\CRIRES). The atmospheric retrieval analysis was published in \citealt{gonzalezpicosESOSupJupSurvey2024} (hereafter GP24) as part of the ESO SupJup survey \citep{regtESOSupJupSurvey2024}.

\subsubsection{Temperature profile}\label{sec:temperature_profile_comparison}
We compare the retrieved temperature profiles in \Cref{fig:PTs} with the results of GP24. While the slope of the profile in the photosphere is similar, we find differences in the photospheric pressure level, likely arising from differences in the retrieved surface gravities and metallicity (see \Cref{fig:logg_metallicity_correlation}). 

GP24 reported a surface gravity ($\logg{} = 3.48 \pm 0.15$) compared to this work ($\logg{} \approx 3.8$--$4.4$). The spread of values presented in this work is attributed to the different wavelength ranges considered (see \Cref{sec:surface_gravity_absolute_abundances}).
The broader contribution function of the JWST/NIRSpec data, enabled by its wide wavelength coverage, provides sensitivity to a larger range of atmospheric pressures. However, pressure regions outside the main contribution function are poorly constrained. The different slopes at the bottom of the atmosphere between \CRIRES and NIRSpec likely reflect differences in spectral coverage and the lack of absolute flux calibration in the \CRIRES data. Uncertainties in flux calibration and its correlation with the brown dwarf radius can also affect the inferred shape of the lower part of the temperature profile.

We also compare our retrieved P-T profiles with Sonora Diamondback models at $T_{\mathrm{eff}}=2400$ K, $\log{g}=4.0$ \citep{morleySonoraSubstellarAtmosphere2024}. The JWST/NIRSpec profiles show good agreement with theoretical models at the peak of the contribution function, but discrepancies are present at higher and lower pressures. At lower pressures, the retrieved temperature profiles are hotter than the theoretical models, which may suggest the presence of clouds such as Al$_2$O$_3$ or Mg$_2$SiO$_4$ condensates \citep{morleySonoraSubstellarAtmosphere2024}. Discrepancies at higher pressures may originate from non-adiabatic effects \citep{tremblinThermocompositionalDiabaticConvection2019} or departures from chemical equilibrium \citep{mukherjeeSonoraSubstellarAtmosphere2024}, but we do not investigate these effects in detail in the present work.

We report the photospheric temperature for each dataset as the temperature within the region encompassing the 90th percentile of the integrated contribution function (see \Cref{fig:PTs}). The photospheric temperatures of TWA 28 are consistent between datasets: $T_{\mathrm{JWST/NIRSpec}} = 2438 \pm 139$ K and $T_{\mathrm{CRIRES}} = 2334 \pm 95$ K. The larger uncertainty for JWST/NIRSpec reflects its broader contribution function. For reference, the derived photospheric temperature of TWA 27A is $T_{\mathrm{TWA 27A}} = 2540 \pm 135$ K, consistent with TWA 27A being marginally hotter than TWA 28.

\subsubsection{Carbon-to-oxygen ratio}\label{sec:C_O_ratio_comparison}
The C/O ratio derived from the G235H grating is $0.60 \pm 0.03$. This value agrees well with the measurement of $0.61 \pm 0.02$ from GP24. We consider the HRS measurement at 1.90-2.45 \micron{} to be the most sensitive probe of the C/O ratio. When including the G395H grating, we find systematically lower C/O ratios. This difference arises from the CO slab emission in the circumstellar disc, which affects the CO fundamental band region. The emission lines effectively decrease the line depth of the photospheric absorption lines, leading to an underestimation of the $^{12}$CO abundance in the atmospheric model.

\subsubsection{Isotope ratios}\label{sec:C_isotope_ratio_comparison}
Measurements of CO isotopologues from the fundamental band provide more precise constraints on the $^{12}$C/$^{13}$C ratio, resulting in narrower posterior distributions (see \Cref{fig:carbon_oxygen_isotope_ratios_calibrated}; see also \citealt{gandhiJWSTMeasurements13C2023a}). However, the accuracy of the $^{12}$C/$^{13}$C ratio retrieved from NIRSpec is a priori uncertain due to potential contamination by the CO slab emission.

We find good agreement between the $^{12}$C/$^{13}$C ratio measured in this work ($75 \pm 2$) and the value reported in GP24 ($81^{+28}_{-19}$). This agreement suggests that slab emission does not significantly affect the derived $^{12}$C/$^{13}$C ratio but we caution that a slight offset to lower values might still be present in the retrieved value due to the potentially underestimated atmospheric \twelveCO{} abundance. For the $^{16}$O/$^{18}$O ratio, GP24 reported a tentative detection of H$_2$$^{18}$O with a value of $205^{+140}_{-62}$. In contrast, our NIRSpec observations result in a significantly higher value of $663^{+53}_{-47}$ for TWA 28. This likely represents a more accurate measurement, as it benefits from both the increased wavelength coverage of NIRSpec and the consistent $^{16}$O/$^{18}$O ratios derived from H$_2$O and CO (see right column of \Cref{fig:carbon_oxygen_isotope_ratios_calibrated}).

We note that measuring \Oratio{} from $K$-band spectroscopy requires very high signal-to-noise ratios to achieve comparable precisions to NIRSpec observations \citep{xuanAreThesePlanets2024}. While ground-based \textit{M}-band spectroscopy may enable measurements of \Oratio{} in the atmospheres of amenable cool stars and brown dwarfs \citep{crossfieldUnusualIsotopicAbundances2019}, current instrumentation may not provide the necessary signal-to-noise at the required spectral resolution to robustly constrain \Oratio{} in the atmospheres of fainter objects. Therefore, JWST/NIRSpec remains a state-of-the-art tool for studying carbon and oxygen isotopic ratios in cool atmospheres.

\subsection{Fundamental parameters}\label{sec:radius_gravity_mass}
We report the fundamental parameters of the two targets as retrieved in this work and compared with literature values in \Cref{tab:physical_properties}. Our retrievals directly fit for the radius and the surface gravity, from which we infer the mass. We estimate the effective temperature of our models as calculated from the integrated flux over a wide wavelength range from 0.35 to 28 \micron{}. We advise to consider these values with caution, as the wavelength range considered here does not cover the entire photosphere of the brown dwarfs and the uncertainties on the effective temperature are likely underestimated.

\subsubsection{Comparison with stellar evolution models}
Stellar evolution models provide powerful tools for estimating the fundamental parameters of stars, brown dwarfs and planets. The cooling tracks of \citet{phillipsNewSetAtmosphere2020} describe the temporal evolution of the radius for objects of different masses. From our flux-calibrated near-infrared spectra, we adopt radii of $R_{\mathrm{TWA\ 28}} = 3.08 \pm 0.03\,R_\mathrm{J}$ and $R_{\mathrm{TWA\ 27A}} = 3.09 \pm 0.02\,R_\mathrm{J}$ for TWA 28 and TWA 27A, respectively, where the uncertainties account for the spread of values between different wavelength ranges.

\begin{figure}[h]
    \centering
    \includegraphics[width=\columnwidth]{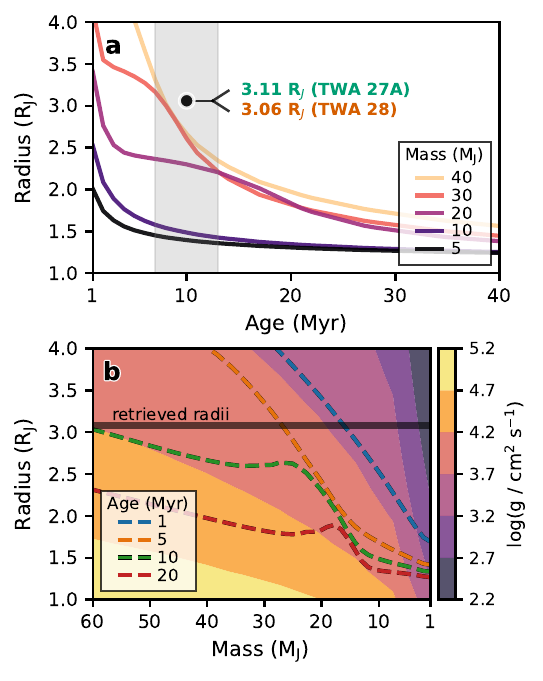}
    \caption{Evolution tracks of young substellar objects. \textbf{a}. Radius tracks for objects with masses between 5 and 40 M$_\mathrm{Jup}$ for the first 40 Myr after formation \citep{phillipsNewSetAtmosphere2020}. \textbf{b}. Physical parameter space of surface gravity for radii between 1 and 4 R$_\mathrm{Jup}$ and masses between 1 and 60 M$_\mathrm{Jup}$. Tracks of different ages are overlaid. The horizontal line indicates the range of retrieved radii for TWA 27A and TWA 28 (3.06-3.11 R$_\mathrm{Jup}$).}
    \label{fig:combined_ybd_properties}
\end{figure}

These values exceed those predicted for objects with masses between 20 and 40~$M_\mathrm{Jup}$ at ages of $\sim$10~Myr (\Cref{fig:combined_ybd_properties}), potentially indicating more massive or younger objects. The striking difference between the full wavelength range (0.97-5.27~\micron{}) and the redder range (1.63-5.27~\micron{}) results deserves particular attention. The 1.63-5.27~\micron{} analysis produces mass estimates of 16.9±1.4 and 28.6±3.3~$M_{\rm J}$ for TWA 27A and TWA 28, respectively, which align well with evolutionary models and previous literature. In contrast, the full wavelength range yields unphysically high masses of 80.5±6.6 and 105.7±5.4~$M_{\rm J}$, suggesting that the bluest spectral region (G140H) introduces systematic biases that affect the retrieved surface gravity and consequently the inferred masses (see \Cref{tab:physical_properties}). Since objects with discs are expected to be younger than those without, the detection of discs in both objects may suggest they are younger than their disc-free counterparts in TWA, as proposed by \citet{venutiXshooterSpectroscopyYoung2019}. Following the 10~Myr evolutionary track (\Cref{fig:combined_ybd_properties}a), predicted surface gravities for objects with masses between 10 and 30~$M_\mathrm{Jup}$ cover the range of $\logg{} \approx 3.7$--$4.2$. The inferred effective temperatures are consistent with the values from the evolutionary models, in particular they agree within one sigma with \citet{venutiXshooterSpectroscopyYoung2019} and \citet{manjavacasMediumresolution09753Mm2024} but are slightly higher (around 200 K) than the values from \citet{cooperUltracoolSpectroscopicOutliers2024}.

Several important caveats must be considered when interpreting these results. The retrieved radii are degenerate with the temperature profile, and may be affected by flux calibration uncertainties that introduce systematic errors not fully accounted for in our analysis. We note that an improved and well-tested flux calibration of NIRSpec data is necessary to obtain reliable results of derived effective temperatures, masses and overall comparison with evolutionary models. Additionally, the relatively low temperatures inferred for the lower atmosphere may indicate an overestimated radius. Previous JWST/NIRSpec analyses by M24 reported lower radii: $R_{\mathrm{TWA\ 28}}$(M24) $= 2.41$--$3.14\,R_\mathrm{J}$ and $R_{\mathrm{TWA\ 27A}}$(M24) $= 2.50$--$2.91\,R_\mathrm{J}$, which align more closely with evolutionary model predictions while remaining broadly consistent with our results.

\subsubsection{Literature mass estimates}
Given that both YBDs belong to the TWA moving group with an estimated age of 10~Myr \citep[e.g.][]{mamajekMovingClusterDistance2005,bellSelfconsistentAbsoluteIsochronal2015}, their masses can be estimated from bolometric luminosity or radius measurements. However, in the absence of dynamical mass constraints, evolutionary model estimates carry significant uncertainties and are inherently model-dependent. 

\begin{table}
    \centering
    \caption{Physical properties of TWA 27A and TWA 28 from different studies.}
    \label{tab:physical_properties}
    \renewcommand{\arraystretch}{1.2}
    \setlength{\tabcolsep}{4pt}
    \begin{tabular}{lcccc}
    \hline
    Target & $T_{\rm eff}$ & Radius & $\log g$ & Mass \\
     & (K) & ($R_{\mathrm{Jup}}$) & (cgs) & ($M_{\mathrm{Jup}}$) \\
    \hline
    \multicolumn{5}{c}{\textit{Venuti et al.} (2019)} \\
    \hline
    TWA 27A & $2640 \pm 20$ & $3.41$ & $3.75 \pm 0.14$ & $19.9 \pm 6.3$ \\
    TWA 28 & $2660 \pm 70$ & $2.82$ & $4.10 \pm 0.30$ & $21.0 \pm 5.2$ \\
    \hline
    \multicolumn{5}{c}{\textit{Manjavacas et al.} (2024)} \\
    \hline
    \multicolumn{5}{l}{\quad ATMO} \\
    TWA 27A & $2600 \pm 100$ & $2.70 \pm 0.21$ & $4.0 \pm 0.5$ & $28.1^{+61.8}_{-19.8}$ \\
    TWA 28 & $2400 \pm 100$ & $2.90 \pm 0.24$ & $4.0 \pm 0.5$ & $32.0^{+67.0}_{-22.2}$ \\
    \multicolumn{5}{l}{\quad BT-Settl} \\
    TWA 27A & $2605 \pm 100$ & $2.70 \pm 0.20$ & $4.0 \pm 0.5$ & $29.3^{+62.8}_{-19.8}$ \\
    TWA 28 & $2577 \pm 100$ & $2.55 \pm 0.19$ & $4.0 \pm 0.5$ & $26.0^{+53.5}_{-16.9}$ \\
    \hline
    \multicolumn{5}{c}{This work} \\
    \hline
    \multicolumn{5}{l}{\quad 0.97-5.27 $\mu$m} \\
    TWA 27A & $2692 \pm 3$ & $3.08 \pm 0.01$ & $4.32 \pm 0.03$ & $80.5^{+6.6}_{-5.6}$ \\
    TWA 28 & $2594 \pm 2$ & $3.10 \pm 0.01$ & $4.44 \pm 0.02$ & $105.7^{+5.4}_{-4.3}$ \\
    \multicolumn{5}{l}{\quad 1.63-5.27 $\mu$m} \\
    TWA 27A & -- & $3.11 \pm 0.01$ & $3.64 \pm 0.04$ & $16.9^{+1.4}_{-1.2}$ \\
    TWA 28 & -- & $3.06 \pm 0.01$ & $3.88 \pm 0.05$ & $28.6^{+3.3}_{-2.3}$ \\
    \hline
    \end{tabular}
    \tablefoot{
    Effective temperature
    (T$_{\rm eff}$) and mass are inferred from the models, while radius and surface gravity are directly retrieved as free parameters.
    ATMO and BT-Settl refer to different atmospheric model grids from M24. 
    Wavelength ranges indicate the spectral coverage: 0.97--5.27~$\mu$m includes all three 
    NIRSpec gratings (G140H, G235H, G395H), while 1.63--5.27~$\mu$m uses only G235H and G395H. }
\end{table}

\cite{venutiXshooterSpectroscopyYoung2019} estimated masses for TWA objects using the \cite{baraffeNewEvolutionaryModels2015} evolutionary tracks based on effective temperature and luminosity. They derived masses of 20 and 21~$M_\mathrm{Jup}$ for TWA 27A and TWA 28, respectively, with typical uncertainties of 5~$M_\mathrm{Jup}$, and corresponding radii of 3.4 and 2.82~$R_\mathrm{J}$. The discrepancy between these radius estimates and our values may reflect degeneracies with disc properties, and the different wavelength coverage of the instrument (0.39--2.45~\micron{}; \citealt{vernetXshooterNewWide2011}).

In contrast, M24 performed spectroscopic analysis of the same JWST/NIRSpec dataset using self-consistent atmospheric models but with a different data reduction. Their analysis produced mass estimates spanning an extremely broad range from low-mass stars to planetary-mass objects (10--90~$M_\mathrm{Jup}$) and surface gravities of $\logg{} = 4.0 \pm 0.5$ for both objects. These broad uncertainties highlight the fundamental challenges of constraining precise masses and surface gravities of low-mass objects from medium-resolution spectroscopy.

\section{Conclusions}\label{sec:conclusions}
JWST/NIRSpec spectroscopy enables unprecedented characterisation of atmospheric and disc properties in young brown dwarfs. Using radiative transfer models, flexible chemistry, and Bayesian retrievals, we report the following findings:
\begin{itemize}
    \item Detection of more than twenty molecular and atomic species, including \twelveCO{}, \water{}, \COtwo{}, SiO, and hydrides (FeH, NaH, CrH, AlH). JWST's broad wavelength coverage and sensitivity, combined with state-of-the-art atmospheric models, reveal species not previously detected in brown dwarf atmospheres such as CO$_2$ and SiO.

    \item Robust detection of \thirteenCO{} and \CeighteenO{} through observations of the CO fundamental band. We derive carbon isotope ratios of $75 \pm 2$ (TWA 28) and $79 ^{+14}_{-11}$ (TWA 27A, after calibration for disc contamination effects). Oxygen isotope ratios are consistent between CO and water molecules in TWA 28, but require correction in TWA 27A due to inaccuracies in disc line emission affecting the inferred CO abundances. We report oxygen isotope measurements of $645^{+80}_{-70}$ and $681^{+53}_{-50}$ for TWA 27A and TWA 28, respectively, derived from water isotopologues.

    \item Well-constrained temperature profiles for both objects, with TWA 28 being marginally cooler than TWA 27A. The photospheric temperature of TWA 28 agrees with previous high-resolution spectroscopy studies, though differences in the temperature profile shape arise from degeneracies between surface gravity and metallicity (see \Cref{fig:PTs}).

    \item Significant excess infrared emission detected in both objects, modelled as warm blackbody rings (approximately 640-650~K) with radii around 12–14~\Rjup{}. We also identify optically thin CO line emission from hot disc gas, which is essential for accurately reproducing the observed spectra at 4.6~\micron{} (see \Cref{fig:ring_emission}) and ensuring consistent fitting of the 2.3~\micron{} CO absorption band.

    \item Carbon-to-oxygen ratios for both objects are consistent with solar values. We infer marginally super-solar metallicities, though the absolute values remain degenerate with surface gravity.

    \item Detailed comparison with previous ground-based high-resolution spectroscopy observations of TWA 28 shows that the JWST/NIRSpec carbon-to-oxygen ratio agrees with the high-resolution spectroscopy value when excluding the region affected by CO slab emission. The retrieved carbon isotope ratios are consistent with high-resolution spectroscopy measurements, suggesting that slab emission does not significantly affect the derived \Cratio{} of TWA 28.

    \item Mass estimates from radii and surface gravities indicate that the blue spectral region (grating G140H) biases surface gravities to unphysically high values. However, the joint fit of G235H and G395H gratings yields mass estimates broadly consistent with evolutionary models and existing literature, supporting mass ranges from 15 to 30 \Mjup.
\end{itemize}

These observations demonstrate JWST/NIRSpec's capabilities for young substellar characterisation. Simultaneous constraints on temperature, chemistry, isotopic ratios, and disc properties enable disentanglement of atmospheric and circumstellar features. CO emission from hot disc gas demonstrates why coupled atmosphere-disc models are essential for accurate abundance measurements in young systems. Future high-resolution spectroscopy may resolve blended spectral features \citep{brandlMETISMidinfraredELT2021,obergObservingCircumplanetaryDisks2023}, but JWST's broad coverage remains crucial for comprehensive analyses. Expanding such studies to younger, low-mass objects will bridge the gap to circumstellar and circumplanetary environments at early stages of evolution.

\begin{acknowledgements}
    We thank E. Manjavacas for providing the original version of the reduced data. We thank G. Giardino for providing calibration measurements of JWST/NIRSpec/IFU used to assess possible instrumental effects. 
     D.G.P and I.S. acknowledge NWO grant OCENW.M.21.010. Support for this work was provided by the NL-NWO Spinoza (SPI.2022.004). This work used the Dutch national e-infrastructure with the support of the SURF Cooperative using grant no. EINF-4556.
    \newline
    \textit{Software}: NumPy \citep{harrisArrayProgrammingNumPy2020}, SciPy \citep{virtanenSciPyFundamentalAlgorithms2020}, Matplotlib \citep{hunterMatplotlib2DGraphics2007}, jwst \citep{bushouseJWSTCalibrationPipeline2025}, petitRADTRANS \citep{mollierePetitRADTRANSPythonRadiative2019}, fastchem \citep{kitzmannFastchemCondEquilibrium2023}, PyAstronomy \citep{czeslaPyAPythonAstronomyrelated2019}, Astropy \citep{collaborationAstropyProjectSustaining2022}, corner \citep{foreman-mackeyCornerPyScatterplot2016}, ExoMol \citep{tennyson2024ReleaseExoMol2024}, HITEMP \citep{rothmanHITEMPHightemperatureMolecular2010}, iris \citep{munoz-romeroJWSTMIRISpectroscopyWarm2024}, pyROX \citep{regtpyROX}
\end{acknowledgements}

\bibliography{library_adstex}

\clearpage
\onecolumn
\appendix

\section{Opacity sources}\label{sec:opacity_sources_appendix}
\begin{table}[h!]
    \centering
    \caption{Opacity sources and references used in this work.}
    \begin{tabular}{cc}
    \hline
    Species & References \\ \hline
    \multicolumn{2}{c}{Line} \\ \hline
    CO  & \cite{rothmanHITEMPHightemperatureMolecular2010}, \cite{liROVIBRATIONALLINELISTS2015} \\
    H$_2$$^{16}$O & \cite{polyanskyExoMolMolecularLine2018} \\
    H$_2$$^{18}$O & \cite{polyanskyExoMolMolecularLine2017} \\
    SiO & \cite{yurchenkoExoMolLineLists2022} \\
    CO$_2$ & \cite{yurchenkoExoMolLineLists2020} \\
    C$_2$H$_2$ & \cite{chubbExoMolMolecularLine2020} \\
    HF & \cite{wilzewskiH2HeCO22016} \\
    VO & \cite{bowesmanExoMolLineLists2024} \\
    TiO & \cite{mckemmishHybridApproachGenerating2024} \\
    OH & \cite{mitevPredissociationDynamicsHydroxyl2024} \\
    H$_2$S & \cite{azzamExoMolMolecularLine2016}, \cite{chubbMarvelAnalysisMeasured2018} \\
    HCl & \cite{gordonHITRAN2016MolecularSpectroscopic2017} \\
    FeH & \cite{dulickLineIntensitiesMolecular2003} \\
    TiH & \cite{burrowsSpectroscopicConstantsAbundances2005}, \cite{bernathMoLLISTMolecularLine2020} \\
    AlH & \cite{yurchenkoExoMolLineLists2024} \\
    Na,K & \cite{allardNewStudyLine2019a} \\
    Fe, Ca, Ti, Mg & \cite{castelliNewGridsATLAS92003} \\\hline
    \multicolumn{2}{c}{Rayleigh scattering} \\ \hline
    H$_2$ & \cite{dalgarnoRayleighScatteringMolecular1962} \\
    He & \cite{chanRefractiveIndexHelium1965} \\ \hline
    \multicolumn{2}{c}{Quasi-continuum} \\ \hline
    H$_2$-H$_2$ CIA & \cite{borysowCollisioninducedRototranslationalAbsorption1988} \\
    H$_2$-He CIA & \cite{borysowCollisioninducedRototranslationalAbsorption1988} \\
    H$^-$ & \cite{grayObservationAnalysisStellar2022} \\
    \end{tabular}
    \label{tab:opacity_references}
\end{table}
\clearpage
\section{Best fit model spectra}\label{sec:best_fit_model_spectra}
\begin{figure}[h]
    \centering
    \includegraphics[width=\textwidth]{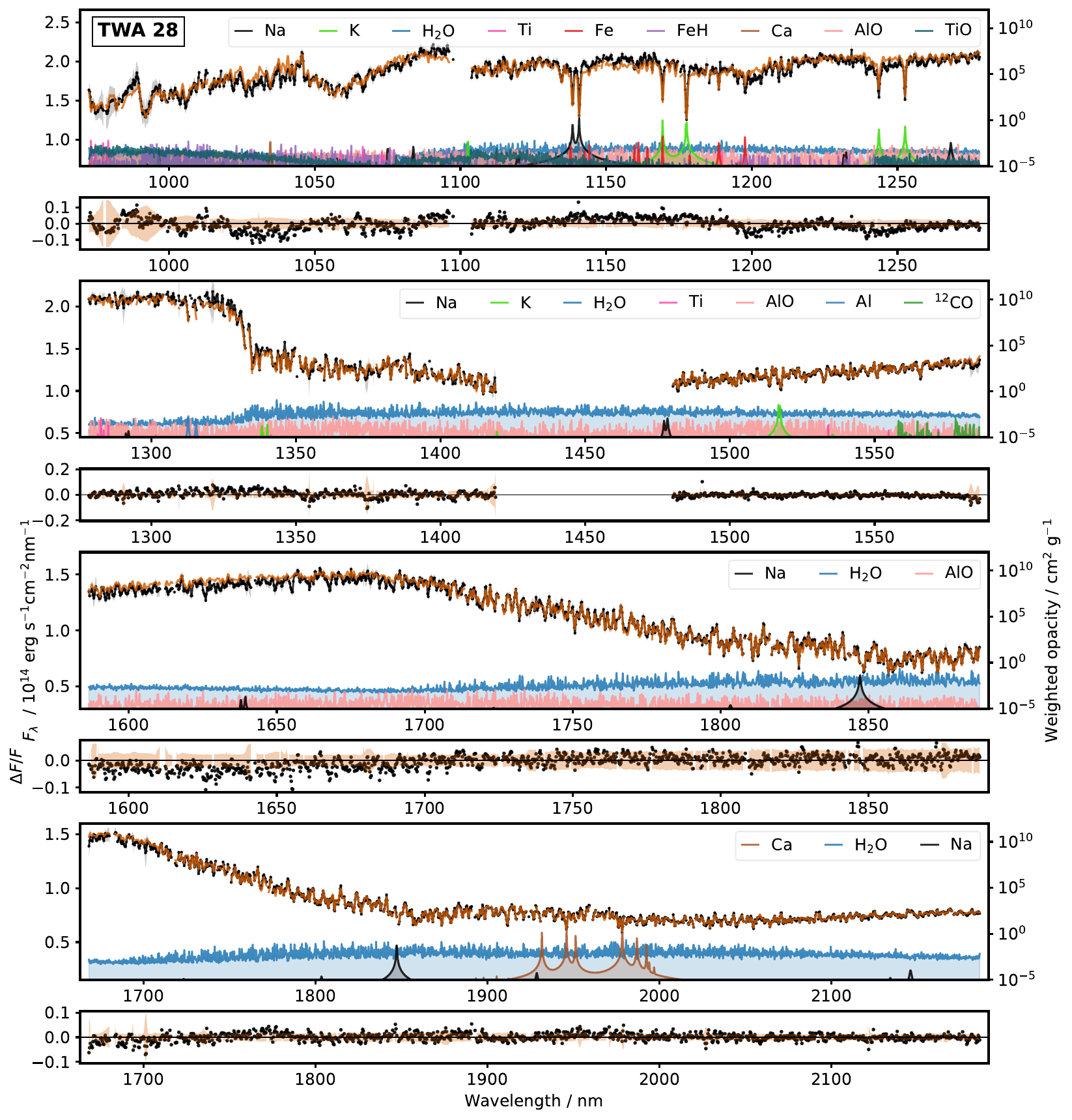}
    \caption{Molecular opacity contributions to the best fit model. The observed spectra (black) and the best fit model spectrum of TWA 28 (orange) are shown with the opacity sources sorted by contribution in each segment of the spectrum. The smaller panels of each segment show the relative residuals as defined in \Cref{fig:full_spec}. The opacity is weighted by the retrieved abundances of the species to scale the contribution of each species. The observed spectra and best fit models are available at \citealt{gonzalezpicosDisentanglingDiscAtmospheric2025} (\url{doi.org/10.5281/zenodo.15781138}).}
    \label{fig:spec_opacities_1}
\end{figure}

\begin{figure}[h]
    \centering
    \includegraphics[width=\textwidth]{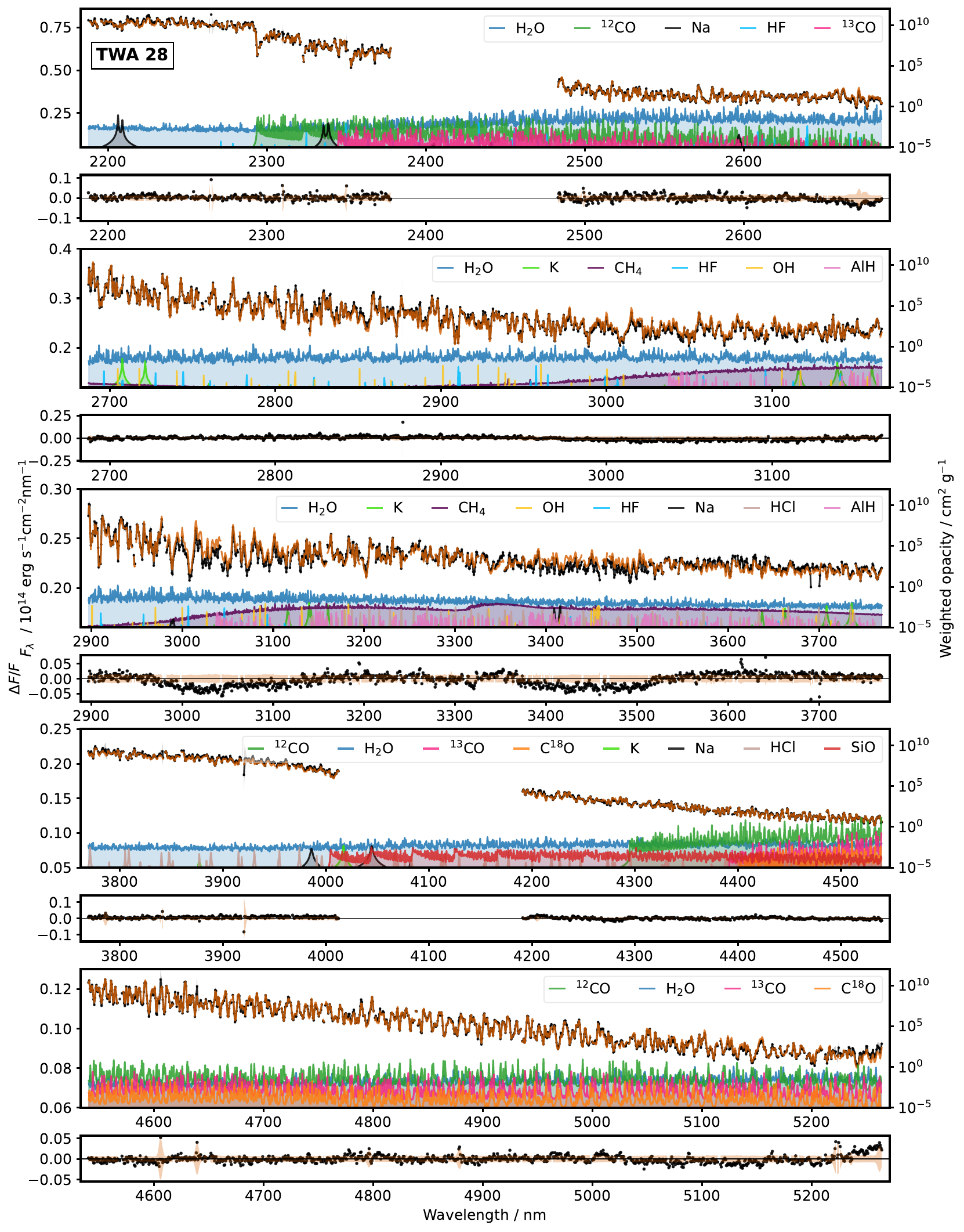}
    \label{fig:spec_opacities_2}
    \captionsetup{labelformat=empty}
    \caption*{Figure B.1 (continued)}
\end{figure}

\begin{figure}[h]
    \centering
    \includegraphics[width=\textwidth]{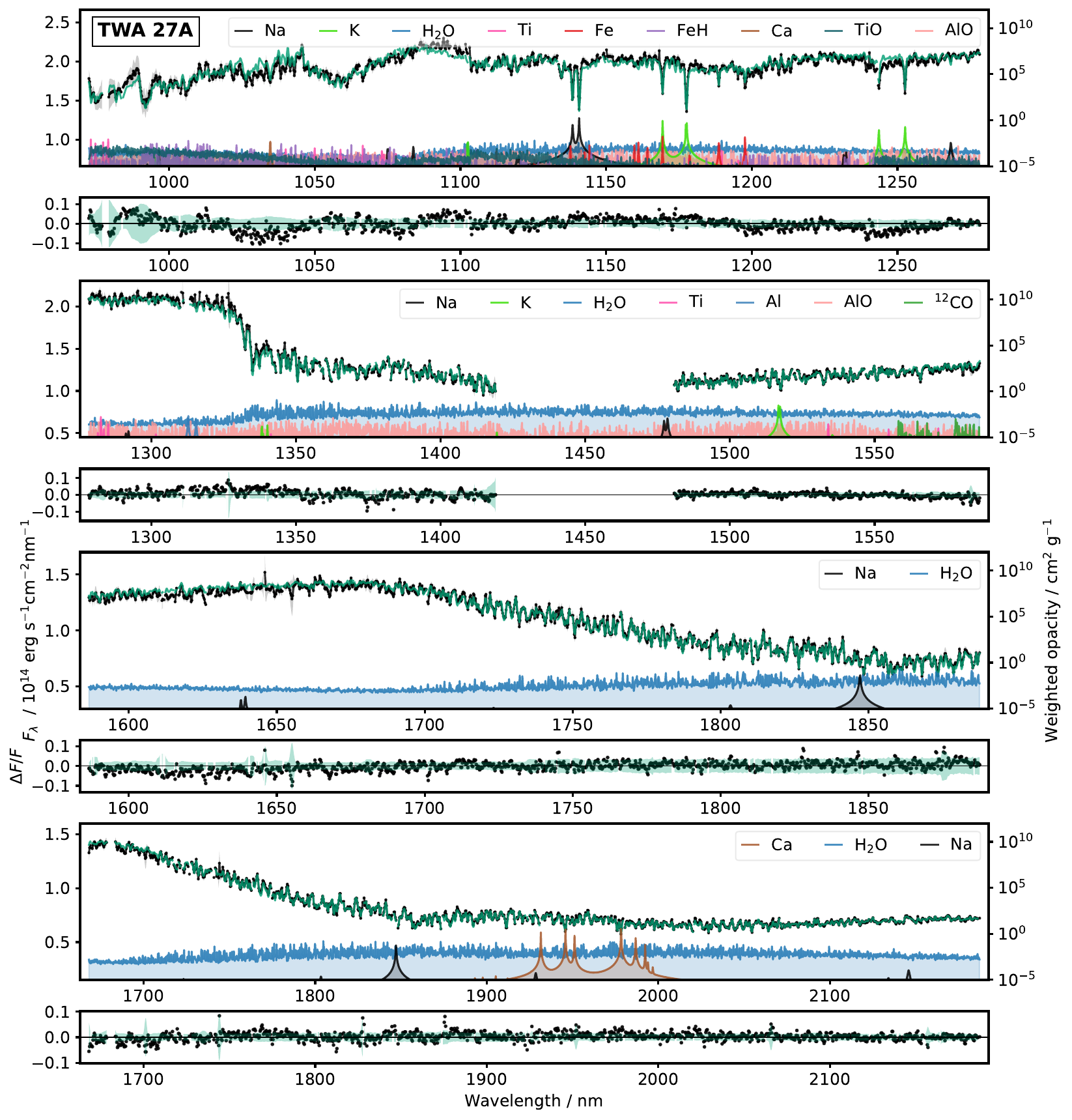}
    \caption{Same as \Cref{fig:spec_opacities_1} but for TWA 27A.}
    \label{fig:spec_opacities_TWA27A}
\end{figure}
\begin{figure}[h]
    \centering
    \includegraphics[width=\textwidth]{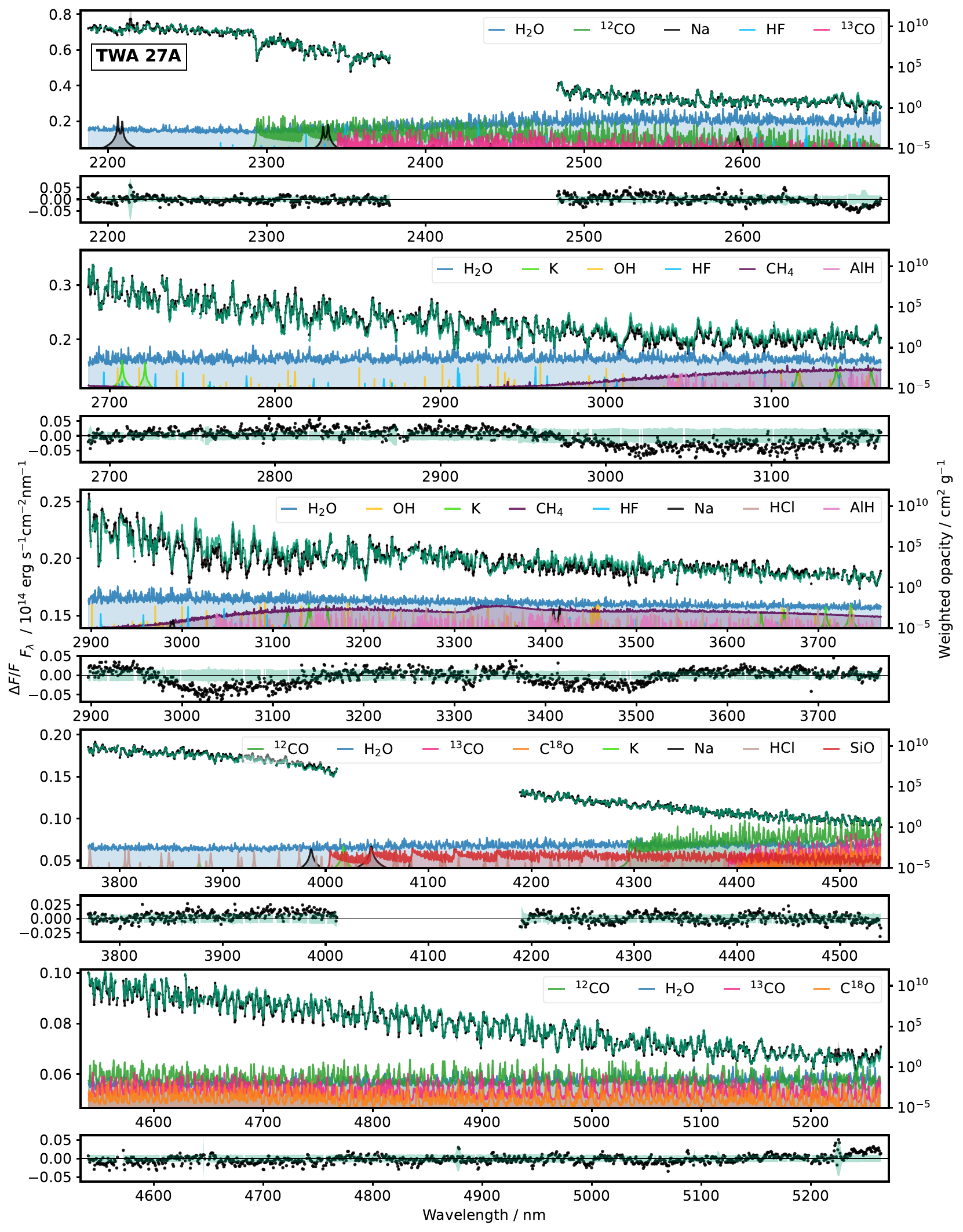}
    \label{fig:spec_opacities_TWA27A_2}
    \captionsetup{labelformat=empty}
    \caption*{Figure B.2 (continued)}
\end{figure}
\clearpage

\section{Best fit model parameters}\label{sec:best_fit_model_parameters}
\begin{table*}[h!]
    \centering
    
    \caption{Summary of the free parameters and the retrieved values with 1$\sigma$ uncertainties. The prior ranges and the distributions used (uniform or normal) are indicated.}
    \renewcommand{\arraystretch}{1.5}
    \begin{tabular}{lllll}
    \hline
     Parameter                                            & Description                                                           & Prior Range               & TWA27A                     & TWA28                      \\
    \hline
     $R_\mathrm{p}$                                       & Radius in Jupiter radii                                               & $\mathcal{U}$(2.2, 3.8)   & $+3.084^{+0.005}_{-0.005}$ & $+3.096^{+0.003}_{-0.004}$ \\
     $v_{\text{rad}} / \text{km s}^{-1}$                  & Radial velocity                                                       & $\mathcal{U}$(-30, 30)    & $+7.98^{+0.12}_{-0.12}$    & $+11.26^{+0.07}_{-0.07}$   \\
     $\log\ g$                                            & Surface gravity of the atmosphere                                     & $\mathcal{U}$(3.0, 4.5)   & $+4.32^{+0.03}_{-0.03}$    & $+4.44^{+0.02}_{-0.02}$    \\
     $T_0$                                                & Temperature at the bottom of the atmosphere                           & $\mathcal{U}$(3000, 8000) & $+3496^{+34}_{-23}$        & $+3367^{+14}_{-12}$        \\
     $\log P_{\text{RCE}} / \text{bar}$                   & Pressure at radiative-convective equilibrium                          & $\mathcal{U}$(-3.0, 1.0)  & $-1.15^{+0.05}_{-0.05}$    & $-0.75^{+0.02}_{-0.04}$    \\
     $\log \Delta P_{\text{low}} / \text{bar}$            & Pressure spacing for lower atmosphere layers                          & $\mathcal{U}$(0.2, 1.6)   & $+0.47^{+0.04}_{-0.03}$    & $+0.24^{+0.03}_{-0.02}$    \\
     $\log \Delta P_{\text{high}} / \text{bar}$           & Pressure spacing for upper atmosphere layers                          & $\mathcal{U}$(0.2, 1.6)   & $+1.44^{+0.08}_{-0.09}$    & $+1.13^{+0.11}_{-0.10}$    \\
     $\nabla_{T,0}$                                       & Temperature gradient at $P_0 = 100$ bar                               & $\mathcal{U}$(0.04, 0.38) & < +0.040                   & < +0.040                   \\
     $\nabla_{T,1}$                                       & Temperature gradient at $P_1=P_{\text{RCE}}-2 \Delta P_{\text{low}}$  & $\mathcal{U}$(0.04, 0.38) & $+0.044^{+0.003}_{-0.002}$ & $+0.042^{+0.001}_{-0.001}$ \\
     $\nabla_{T,2}$                                       & Temperature gradient at $P_2=P_{\text{RCE}}-1 \Delta P_{\text{low}}$  & $\mathcal{U}$(0.04, 0.38) & $+0.124^{+0.002}_{-0.002}$ & $+0.122^{+0.002}_{-0.002}$ \\
     $\nabla_{T,3}$                                       & Temperature gradient at $P_3=P_{\text{RCE}}+1 \Delta P_{\text{high}}$ & $\mathcal{U}$(0.00, 0.38) & $+0.036^{+0.004}_{-0.004}$ & $+0.072^{+0.004}_{-0.004}$ \\
     $\nabla_{T,4}$                                       & Temperature gradient at $P_4=P_{\text{RCE}}+2 \Delta P_{\text{high}}$ & $\mathcal{U}$(0.00, 0.38) & $+0.121^{+0.005}_{-0.008}$ & $+0.055^{+0.004}_{-0.005}$ \\
     $\nabla_{T,5}$                                       & Temperature gradient at $P_5=10^{-5}$ bar                             & $\mathcal{U}$(0.00, 0.38) & $+0.113^{+0.010}_{-0.015}$ & $+0.006^{+0.005}_{-0.004}$ \\
     $\nabla_{T,RCE}$                                     & Temperature gradient at $P_{\text{RCE}}$                              & $\mathcal{U}$(0.04, 0.38) & $+0.128^{+0.002}_{-0.001}$ & $+0.127^{+0.002}_{-0.002}$ \\
     $\log R^{\text{bb}} / R_{\text{jup}}$                & Effective radius of the blackbody emission                            & $\mathcal{U}$(0, 2)       & $+1.077^{+0.005}_{-0.006}$ & $+1.142^{+0.003}_{-0.003}$ \\
     $T^{\text{bb}}_{\text{eff}} / \text{K}$              & Temperature of the blackbody emission                                 & $\mathcal{U}$(300, 900)   & $+642.91^{+4.02}_{-3.71}$  & $+653.20^{+1.93}_{-1.91}$  \\
     $\log R^{\text{slab}} / R_{\text{jup}}$              & Effective radius of the slab model                                    & $\mathcal{U}$(0, 3)       & $+0.861^{+0.022}_{-0.017}$ & $+1.448^{+0.148}_{-0.122}$ \\
     $\log N^{\text{slab}}_{\text{mol}} / \text{cm}^{-2}$ & Column density of slab model                                          & $\mathcal{U}$(14, 18)     & $+17.33^{+0.06}_{-0.08}$   & $+15.98^{+0.26}_{-0.31}$   \\
     $\log T^{\text{slab}}_{\text{ex}} / \text{K}$        & Excitation temperature of slab model                                  & $\mathcal{U}$(2.7, 3.3)   & > +3.300                   & $+3.227^{+0.009}_{-0.009}$ \\
     $v^{\text{slab}} / \text{km s}^{-1}$                 & Radial velocity of disk emission                                      & $\mathcal{U}$(-60, 60)    & $-0.34^{+1.58}_{-1.60}$    & $+7.62^{+0.90}_{-0.88}$    \\
     $\log l_{\text{G}} / \text{km s}^{-1}$               & Global correlation length                                             & $\mathcal{U}$(1.4, 2.6)   & $+1.901^{+0.016}_{-0.015}$ & $+1.900^{+0.013}_{-0.011}$ \\
     $\log\ b$(g140h)                                     & Error scaling factor for G140H grating                                & $\mathcal{U}$(0, 3)       & $+1.62^{+0.16}_{-0.16}$    & $+1.85^{+0.05}_{-0.05}$    \\
     $\log\ b$(g235h)                                     & Error scaling factor for G235H grating                                & $\mathcal{U}$(0, 3)       & $+1.92^{+0.04}_{-0.05}$    & $+1.88^{+0.02}_{-0.02}$    \\
     $\log\ b$(g395h)                                     & Error scaling factor for G395H grating                                & $\mathcal{U}$(0, 3)       & $+1.17^{+0.15}_{-0.17}$    & $+1.32^{+0.03}_{-0.03}$    \\
     $\alpha_{12CO}$                                      & Deviation from chemical equilibrium for $^{12}$CO                     & $\mathcal{N}$(0, 1)       & $+0.16^{+0.03}_{-0.03}$    & $+0.39^{+0.02}_{-0.02}$    \\
     $\alpha_{H2O}$                                       & Deviation from chemical equilibrium for H$_2$O                        & $\mathcal{N}$(0, 1)       & $+0.43^{+0.03}_{-0.03}$    & $+0.55^{+0.02}_{-0.02}$    \\
     $\alpha_{SiO}$                                       & Deviation from chemical equilibrium for SiO                           & $\mathcal{N}$(0, 1)       & $+0.26^{+0.04}_{-0.04}$    & $+0.43^{+0.03}_{-0.02}$    \\
     $\alpha_{CO2}$                                       & Deviation from chemical equilibrium for CO$_2$                        & $\mathcal{N}$(0, 1)       & $+0.37^{+0.04}_{-0.04}$    & $+0.31^{+0.03}_{-0.03}$    \\
     $\alpha_{TiO}$                                       & Deviation from chemical equilibrium for TiO                           & $\mathcal{N}$(0, 1)       & $-0.17^{+0.04}_{-0.03}$    & $-0.06^{+0.02}_{-0.02}$    \\
     $\alpha_{VO}$                                        & Deviation from chemical equilibrium for VO                            & $\mathcal{N}$(0, 1)       & $+0.30^{+0.04}_{-0.03}$    & $+0.42^{+0.02}_{-0.02}$    \\
     $\alpha_{FeH}$                                       & Deviation from chemical equilibrium for FeH                           & $\mathcal{N}$(0, 1)       & $+0.50^{+0.04}_{-0.04}$    & $+0.77^{+0.02}_{-0.02}$    \\
     $\alpha_{CrH}$                                       & Deviation from chemical equilibrium for CrH                           & $\mathcal{N}$(0, 1)       & $+0.01^{+0.09}_{-0.10}$    & $+0.23^{+0.07}_{-0.07}$    \\
    \hline
    \end{tabular}
    \label{tab:free_params_comparison_0}
\end{table*}
\clearpage

\begin{table*}[h!]
    \centering
    \captionsetup{labelformat=empty}
    \caption*{Table C.1 (continued)}
    \renewcommand{\arraystretch}{1.5}
    \begin{tabular}{lllll}
    \hline
     Parameter                            & Description                                      & Prior Range             & TWA27A                     & TWA28                      \\
    \hline
     $\alpha_{NaH}$                       & Deviation from chemical equilibrium for NaH      & $\mathcal{N}$(0, 1)     & $+0.83^{+0.08}_{-0.08}$    & $+1.16^{+0.05}_{-0.06}$    \\
     $\alpha_{AlH}$                       & Deviation from chemical equilibrium for AlH      & $\mathcal{N}$(0, 1)     & $+0.01^{+0.23}_{-0.31}$    & $+0.43^{+0.06}_{-0.06}$    \\
     $\alpha_{HF}$                        & Deviation from chemical equilibrium for HF       & $\mathcal{N}$(0, 1)     & $+0.08^{+0.16}_{-0.17}$    & $-0.14^{+0.12}_{-0.13}$    \\
     $\alpha_{HCl}$                       & Deviation from chemical equilibrium for HCl      & $\mathcal{N}$(0, 1)     & $-0.19^{+0.09}_{-0.08}$    & $-0.25^{+0.07}_{-0.08}$    \\
     $\alpha_{Na}$                        & Deviation from chemical equilibrium for Na       & $\mathcal{N}$(0, 1)     & $-0.37^{+0.04}_{-0.04}$    & $-0.17^{+0.03}_{-0.03}$    \\
     $\alpha_{Ca}$                        & Deviation from chemical equilibrium for Ca       & $\mathcal{N}$(0, 1)     & $+0.09^{+0.05}_{-0.04}$    & $+0.40^{+0.02}_{-0.03}$    \\
     $\alpha_{K}$                         & Deviation from chemical equilibrium for K        & $\mathcal{N}$(0, 1)     & $-0.28^{+0.04}_{-0.04}$    & $-0.04^{+0.03}_{-0.03}$    \\
     $\alpha_{Mg}$                        & Deviation from chemical equilibrium for Mg       & $\mathcal{N}$(0, 1)     & $-0.23^{+0.38}_{-0.45}$    & $-0.23^{+0.40}_{-0.47}$    \\
     $\alpha_{Al}$                        & Deviation from chemical equilibrium for Al       & $\mathcal{N}$(0, 1)     & $-0.33^{+0.18}_{-0.20}$    & $+0.77^{+0.04}_{-0.05}$    \\
     $\alpha_{Fe}$                        & Deviation from chemical equilibrium for Fe       & $\mathcal{N}$(0, 1)     & $-0.13^{+0.07}_{-0.07}$    & $+0.22^{+0.05}_{-0.06}$    \\
     $\alpha_{Ti}$                        & Deviation from chemical equilibrium for Ti       & $\mathcal{N}$(0, 1)     & $+0.03^{+0.08}_{-0.08}$    & $+0.21^{+0.04}_{-0.05}$    \\
     $\log\ \mathrm{CH_4}$                & Abundance of CH$_4$                              & $\mathcal{U}$(-14, -2)  & $-5.503^{+0.036}_{-0.033}$ & $-5.369^{+0.024}_{-0.022}$ \\
     $\log\ \mathrm{AlO}$                 & Abundance of AlO                                 & $\mathcal{U}$(-14, -2)  & $-7.84^{+0.04}_{-0.04}$    & $-7.65^{+0.04}_{-0.05}$    \\
     $\log \text{H}^{-}$                  & Abundance of H$^{-}$ bound-free opacity          & $\mathcal{U}$(-12, -7)  & $-8.94^{+0.04}_{-0.03}$    & $-8.78^{+0.02}_{-0.02}$    \\
     $\log\ \mathrm{^{12}CO/^{13}CO}$     & Carbon isotope ratio $^{12}$C/$^{13}$C in CO     & $\mathcal{U}$(1.0, 3.0) & $+1.72^{+0.03}_{-0.03}$    & $+1.87^{+0.01}_{-0.01}$    \\
     $\log\ \mathrm{C^{16}O/C^{18}O}$     & Oxygen isotope ratio $^{16}$O/$^{18}$O in CO     & $\mathcal{U}$(1.5, 4.0) & $+2.63^{+0.04}_{-0.04}$    & $+2.84^{+0.03}_{-0.03}$    \\
     $\log\ \mathrm{C^{16}O/C^{17}O}$     & Oxygen isotope ratio $^{16}$O/$^{17}$O in CO     & $\mathcal{U}$(1.5, 4.0) & $+3.27^{+0.06}_{-0.06}$    & $+3.61^{+0.04}_{-0.04}$    \\
     $\log\ \mathrm{H_2^{16}O/H_2^{18}O}$ & Oxygen isotope ratio $^{16}$O/$^{18}$O in H$_2$O & $\mathcal{U}$(1.5, 4.0) & $+2.81^{+0.05}_{-0.05}$    & $+2.83^{+0.03}_{-0.03}$    \\
    \hline
    \end{tabular}
    \label{tab:free_params_comparison_1}
\end{table*}

\section{Vertical composition profiles}\label{sec:volume_mixing_ratios}
\begin{figure}[h!]
    \centering
    \includegraphics[width=\textwidth]{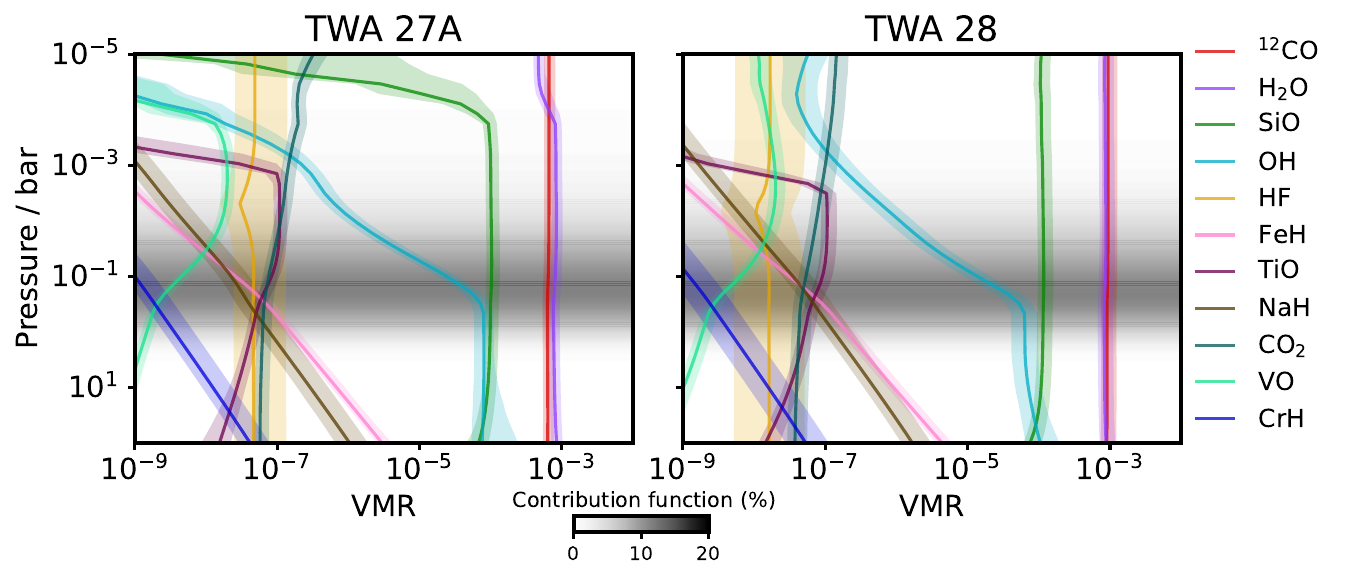}
    \caption{Vertical composition profiles. Volume mixing ratios of the retrieved molecules as a function of pressure for TWA 27A and TWA 28. Species are colour-coded as indicated in the legend. Darker background regions indicate higher values of the contribution function.}
    \label{fig:VMRs}
\end{figure}
    
\clearpage
\section{Extended data with Spitzer/IRS observations}\label{sec:spitzer_data
}
\begin{figure}[h]
    \centering
    \includegraphics[width=\textwidth]{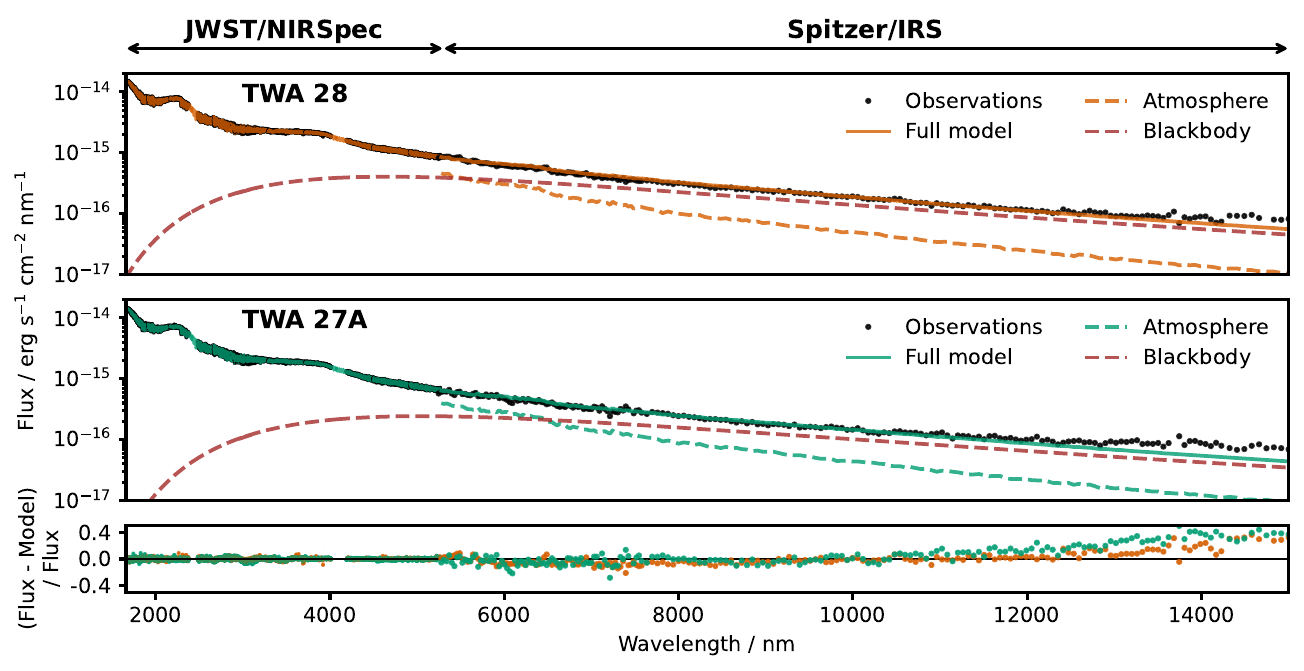}
    \caption{Extended wavelength model validation with Spitzer/IRS data. Best fit model spectra of TWA 27A and TWA 28 extended to the Spitzer/IRS wavelength range \citep{riazNewBrownDwarf2008}. Observed spectra are shown in black and best fit model spectra are shown as solid lines. Dashed brown lines show blackbody emission from the disc and the dashed line with corresponding colour shows the atmospheric model. Data beyond 11 and 12 \micron{} starts to clearly deviate from our best fit model for TWA 27A and TWA 28, respectively.}
    \label{fig:spec_spitzer}
\end{figure}

\end{document}